\documentclass[10pt]{article}
\usepackage{amsmath,graphicx, enumerate,textcomp,amssymb}
\usepackage[letterpaper,top=0.5in, bottom=1.in, left=0.8in, right=0.8in, headsep=0.in, centering]{geometry}
\usepackage{amssymb}
\usepackage{authblk}
\usepackage{float,color}

\newcommand{\be}{\begin{equation}}
\newcommand{\ee}  {\end{equation}}
\newcommand{\bd}{\begin{displaymath}}
\newcommand{\ed}{  \end{displaymath}}

\newcommand{\Lw}{{\rm Le}}

\newcommand{\Le}{\hbox{Le}}

\newcommand{\eps}{\varepsilon}

\newcommand{\tb}{\textcolor{black}}

\makeatletter
\let\@fnsymbol\@arabic
\makeatother

\title{Strongly nonlinear asymptotic model of cellular instabilities in premixed flames with stepwise ignition temperature kinetics}

\author{Nathan Kilker\thanks{\tt npk5@zips.uakron.edu} }
\author{Dmitry Golovaty\thanks{\tt dmitry@uakron.edu} }
\author{Peter  V. Gordon\thanks{\tt pgordon@uakron.edu}}
\affil{Department of  Mathematics, The University of Akron, Akron, Ohio 44325}
\author{Leonid  Kagan\thanks{\tt kaganleo@post.tau.ac.il} }
\author{Gregory I.  Sivashinsky\thanks{\tt grishas@post.tau.ac.il}}
\affil{School of Mathematical Sciences, Tel Aviv University, Tel Aviv 69978, Israel}

\begin{document}

\maketitle


\bigskip


\begin{abstract}\tb{
In this paper we consider ignition-temperature, first-order reaction model of thermo-diffusive combustion that describes dynamics of thick flames
arising, for example,  in a theory of combustion of  hydrogen-oxygen and ethylene-oxygen mixtures. 
These flames often assume the shape of propagating curved interfaces that can be identified with level sets corresponding to a prescribed ignition temperature. The present paper is concerned with the analysis of such interfaces in two spatial
dimensions under a single assumption of their small curvature.}

\tb{We derive a strongly nonlinear evolution equation that governs the dynamics of an interface. This equation relates the normal velocity of the interface to its curvature, the derivatives of the  curvature with respect to the arc-length of the interface and physical parameters of the problem.
We study solutions of the evolution equation for various parameter regimes and discuss the ranges of validity of the corresponding simplified models. Our theoretical findings are illustrated and supported by numerical simulations.}
\end{abstract}

\section{Introduction}
This paper is concerned with the  analysis of dynamics of planar curved traveling interfaces for the thermo-diffusive model of flame propagation with stepwise temperature kinetics and first order reaction. In non-dimensional form, the model reads:
\begin{eqnarray}\label{eq:1}
\Theta_t=\Delta \Theta+W(\Theta,\Phi),
\end{eqnarray}
\begin{eqnarray}\label{eq:2}
\Phi_t=\Lw^{-1}\Delta \Phi-W(\Theta,\Phi),
\end{eqnarray}
\begin{eqnarray}\label{eq:3}
W(\Theta,\Phi)=
\left\{ 
\begin{array}{lll}
A \Phi & \mbox{at} & \Theta\ge \Theta_i, \\
0 & \mbox{at} &\Theta<\Theta_i.
\end{array}
\right.
\end{eqnarray}
Here $\Theta$ and $\Phi$  are appropriately normalized temperature and concentration of deficient reactant, ${\rm Le}>0$ is the Lewis number,
$0<\Theta_i<1$ is the ignition temperature, and $A>0$ is a normalizing factor. Equations \eqref{eq:1} and \eqref{eq:2}
correspond to conservation of energy and a reactive component of premixed fuel/oxidizer, respectively, whereas \eqref{eq:3}
prescribes the reaction rate.

The model \eqref{eq:1}--\eqref{eq:3}  is by no means new and was used by several authors, predominantly due to its simplicity
and mathematical tractability (see, e.g., \cite{i8,BGKS15,i7,i5,i6,i4}). Our interest in \eqref{eq:1}--\eqref{eq:3}, however, is motivated by recent advances in understanding
of overall effective chemical kinetics for hydrogen-oxygen and ethylene-oxygen mixtures.  Indeed, recent theoretical and numerical
studies based on the detailed chemistry mechanisms revealed that the global activation energy $E_g$ for such mixtures appears to be high at low enough temperatures and
low at high enough temperatures \cite{i2,i1,i3}.
These findings strongly suggest the conclusion that an improved description of combustion waves
for  hydrogen-oxygen and ethylene-oxygen mixtures can be achieved by
employing the global one-step kinetics with an
appropriately modified Arrhenius exponent.
 Moreover, to sharpen the physical
picture one may consider the extreme situation where $E_g = \infty$, for temperatures \tb{lower} than some
effective ignition temperature and $E_g = 0$ for \tb{higher} temperatures. The latter observation leads
to model \eqref{eq:1}--\eqref{eq:3}, but now on entirely physical grounds.

\begin{figure}[htb]
\hspace{.5cm}
\begin{minipage}[c]{0.4\textwidth}%
\begin{center}
\includegraphics[height=1.8in]{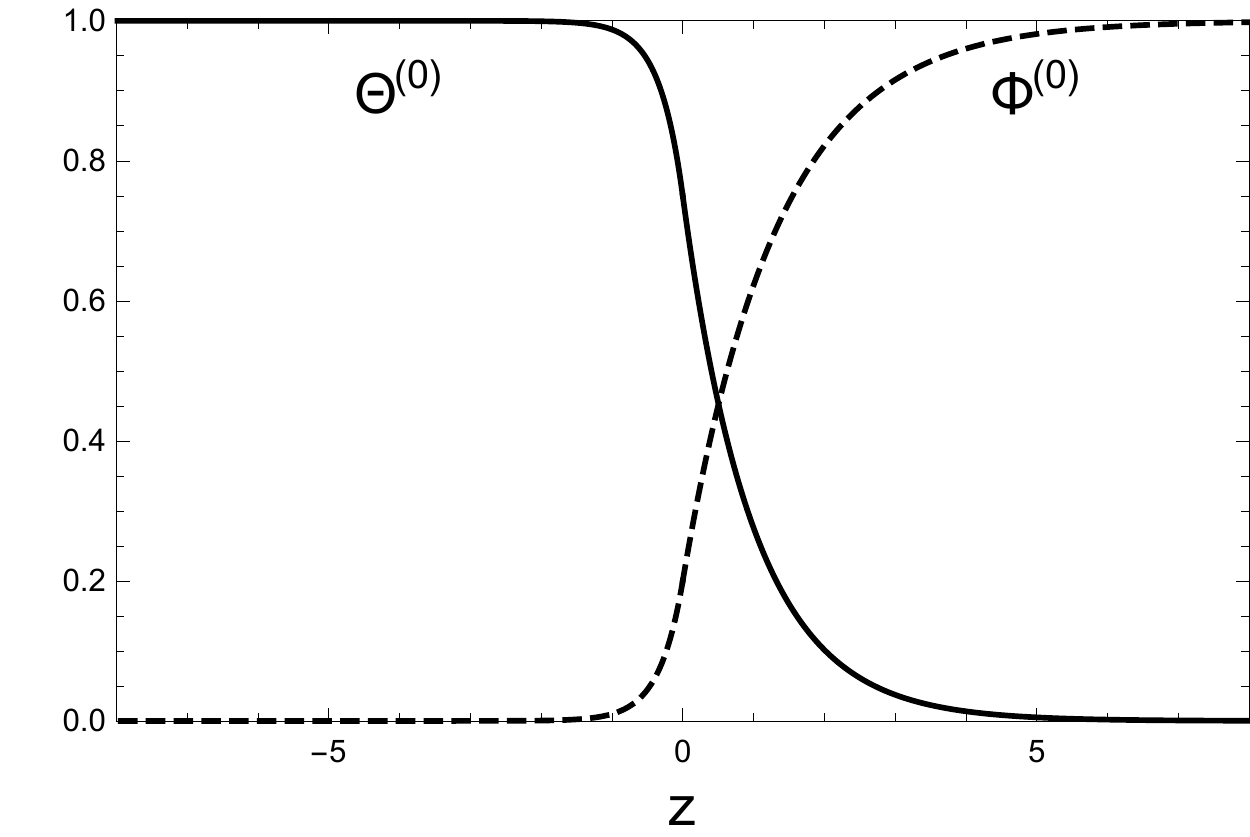}
\par\end{center}%
\end{minipage}\hspace{.8cm} %
\hspace{1cm}
\begin{minipage}[c]{0.4\textwidth}
\includegraphics[height=1.8in]{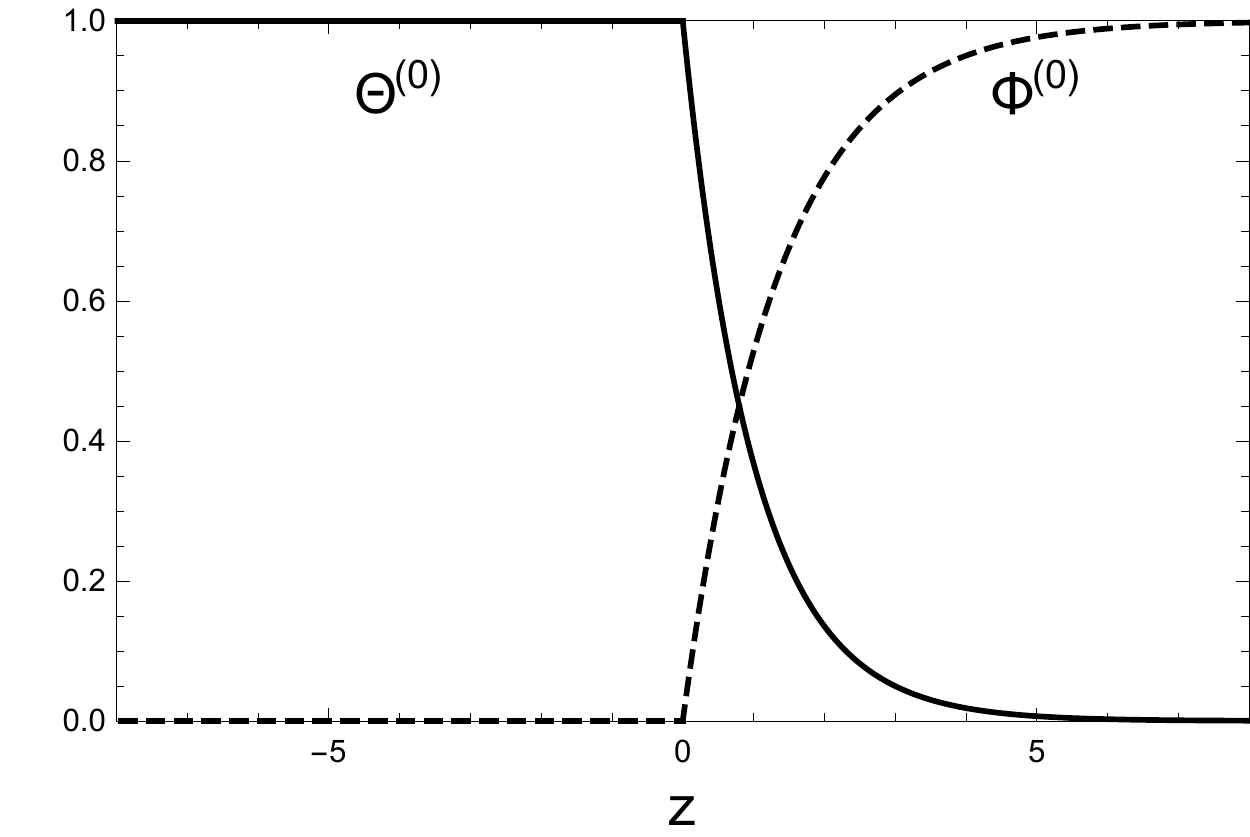}
\end{minipage}
\caption{Spatial profiles of $\Theta^{(0)}(z)$ (solid line) and $\Phi^{(0)}(z)$ (dashed line). \tb{Left panel:  model \eqref{eq:1}--\eqref{eq:3} at $\Theta_i=0.75$ and $\Lw=0.75$;  Right panel: model \eqref{eq:1}--\eqref{eq:2} with \eqref{eq:3} replaced by the $\delta$-function corresponding to the limiting point source kinetics for the conventional Arrhenius law and $\Lw=0.75$.}}
  \label{fig:fr}
\end{figure}

One of the principal features of the model \eqref{eq:1}--\eqref{eq:3}  is that it admits a unique (up to translations), one-dimensional traveling interface
solution. By choosing 
\begin{eqnarray}\label{eq:6}
A=\frac{\Theta_i}{1-\Theta_i}\left( 1+\frac{\Theta_i}{\Lw(1-\Theta_i)}\right),
\end{eqnarray}
we ensure that the interface propagates with the speed one so that both temperature and concentration fields depend on a single variable $z=x-t$. Here $x$ denotes the spatial variable and $t>0$ is time.
The traveling interface solution is explicitly given by the following formula:
\begin{eqnarray}\label{eq:4}
\Theta^{(0)}(z)=\left\{
\begin{array}{ll}
\Theta_i\exp(-z), & z \ge 0, \\
1-(1-\Theta_i)\exp\left(\frac{\Theta_i}{1-\Theta_i} z \right), & z<0,
\end{array}
\right.
\end{eqnarray}
and
\begin{eqnarray}\label{eq:5}
 \Phi^{(0)}(z)=\left\{
\begin{array}{ll}
1-\frac{\Theta_i}{\Theta_i+\Lw(1-\Theta_i)}\exp(-\Lw\,z), & z \ge 0, \\
\frac{\Lw(1-\Theta_i)}{\Theta_i+\Lw(1-\Theta_i)} \exp\left(\frac{\Theta_i}{1-\Theta_i} z \right), & z <0.
\end{array}
\right.
\end{eqnarray}
Typical profiles of the traveling interface solution are depicted in \tb{ the left panel}  in Fig. \ref{fig:fr}.
\tb{The traveling wave solution given by \eqref{eq:4}--\eqref{eq:5} and depicted in the left panel in Fig. \ref{fig:fr} is clearly different from that arising
in conventional thermo-diffusive combustion with the standard Arrhenius kinetics at high
activation energies (depicted in the right panel in the same figure).}  Indeed, in the model
considered here the reaction zone width is of order unity, whereas in the case of Arrhenius kinetics the reaction zone is infinitesimally thin. This fact suggests that it is appropriate to refer to traveling interfaces for stepwise temperature kinetics as thick flames, in contrast to thin flames arising in Arrhenius kinetics.

The presence of one-dimensional  traveling interface solution for the model \eqref{eq:1}--\eqref{eq:3} poses a natural question of interface stability in higher dimensions and, more generally, understanding the dynamics of a curved level set $\Theta=\Theta_i$. From now on, the word {\it interface} will be used to refer to this level set. \tb{ In what follows, we work in two spatial dimensions, with an interface being associated with a curve evolving either in the entire $\mathbb{R}^2$ or in a two-dimensional strip-like domain.}

\begin{figure}[htb]
\centering
\includegraphics[height=2in]{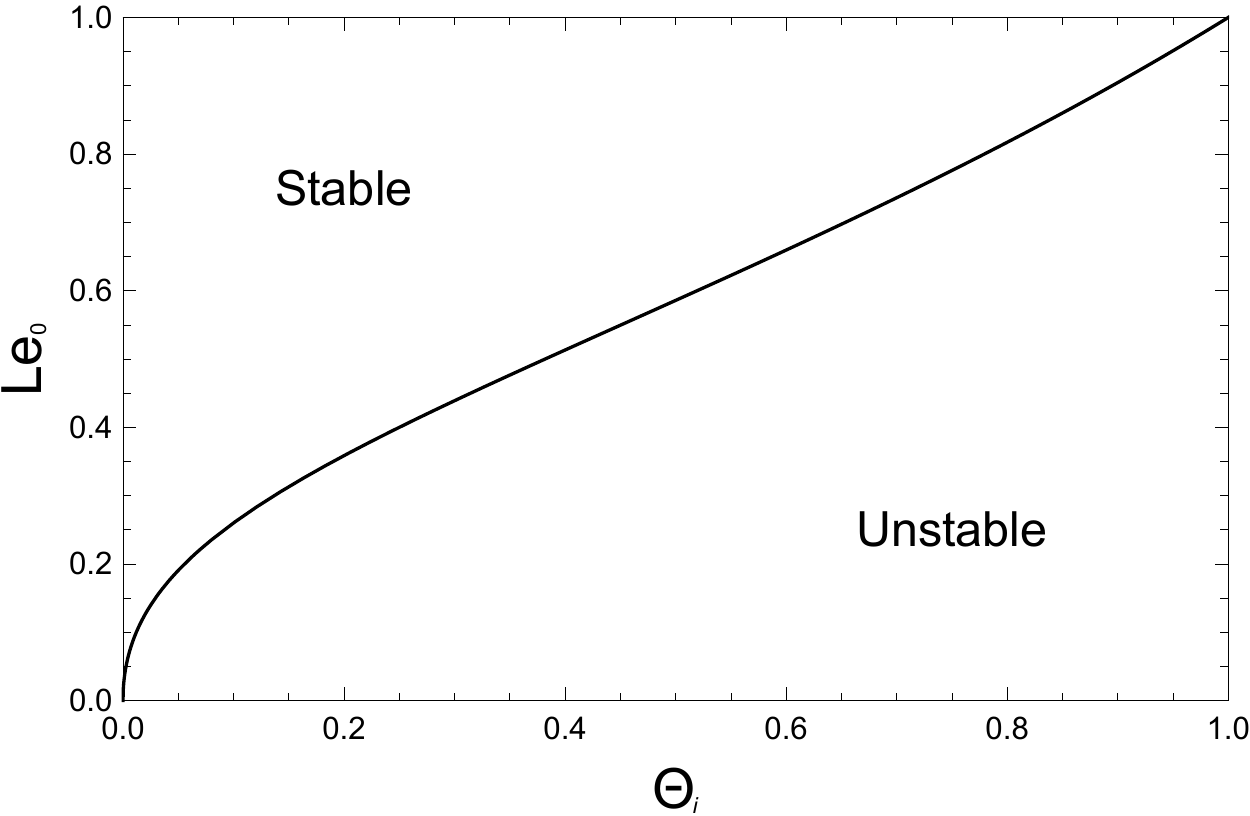}
\caption{Critical Lewis number $\Lw_0$ vs. ignition temperature $\Theta_i$.}
  \label{fig:st}
\end{figure}

In a recent work \cite{BGKS15}, the authors performed the linear stability analysis of planar traveling interface solutions for the system \eqref{eq:1}--\eqref{eq:3}. It was shown that the stability picture depends dramatically on the value of the Lewis number $\Lw$. Specifically, two different instabilities were observed. For $\Lw>1$ the system may exhibit only pulsating instabilities in certain parameter regimes. This type of instability of a traveling interface is characterized by time-periodic oscillations of the mean interface velocity and the shape of the interface. There also exists a critical value of the Lewis number $0<\rm{Le_0}(\Theta_i)<1$,  given by
\begin{equation}
\label{le0}
\mathrm{Le_0}(\Theta_i)=\frac{{\Theta_i}^2 - 3\Theta_i + \left({\Theta_i}^4 + 2 {\Theta_i}^3 - 15 {\Theta_i}^2 + 16\Theta_i\right)^{\frac{1}{2}}}{2 \left({\Theta_i}^2-3 \Theta_i+2\right)},
\end{equation}
such that a planar interface is linearly stable when $\rm{Le_0}<\rm{Le}<1$ and it is {\it cellularly unstable} when $\rm{Le}<\rm{Le_0}$ as shown in Fig. \ref{fig:st}. \tb{We say  that a planar interface is cellularly unstable if it
forms large-scale spatial periodic structures resulting from linear instability in a bounded range of wave numbers with a real instability growth rate \cite{gr_rev}.}

Further, it was established in \cite{BGKS15} that, for Lewis number slightly below critical, i.e.,
\begin{eqnarray}\label{eq:delta}
\delta=\frac{\rm{Le_0}-\rm{Le}}{\rm{Le_0}}\ll 1,
\end{eqnarray}
the perturbed interface appears to be quasi-planar ($k\sim\sqrt{\delta}$) 
and quasi-steady ($\omega\sim \delta^2$). 
Here $\omega$ and $k$ are an instability growth rate and transverse wave number, respectively.
As a result, the dispersion relation reduces to
\begin{eqnarray}\label{eq:disp}
\omega=\delta \lambda k^2-\mu k^4,
\end{eqnarray}
at leading order, where 
\begin{eqnarray}\label{eq:lam}
 \lambda(\Theta_i):= \frac{2\Theta_i^2+\Lw_0(4-\Lw_0)\Theta_i(1-\Theta_i)}{2\Lw_0(1-\Theta_i)[\Theta_i+\Lw_0(1-\Theta_i)]^2},
\end{eqnarray}
 and 
\tb{\begin{eqnarray}\label{eq:mu}
\begin{aligned}
\mu(\Theta_i):= &\frac{(1-\Lw_0)\Theta_i}{4\Lw_0^3[2\Theta_i+(1-\Theta_i)\Lw_0](1-\Theta_i)^2[\Theta_i+(1-\Theta_i)\Lw_0]^2} \\
&\Big\{ 12\Lw_0^2+6\Lw_0^3+2\Lw_0^4+2\Lw_0^5+\Theta_i(24\Lw_0-22\Lw_0^2-17\Lw_0^3-3\Lw_0^4-10\Lw_0^5) \\
&+\Theta_i^2(16-36\Lw_0+4\Lw_0^2+13\Lw_0^3-12\Lw_0^4+19\Lw_0^5)  \\
&-\Theta_i^3(8-12\Lw_0-2\Lw_0^2+13\Lw_0^3-28\Lw_0^4+17\Lw_0^5)  \\
& -\Theta_i^4(4\Lw_0^2-13\Lw_0^3+18\Lw_0^4-7\Lw_0^5)-\Theta_i^5(2\Lw_0^3-3\Lw_0^4+\Lw_0^5)\Big\}.
\end{aligned}
\end{eqnarray}}
Both $\lambda(\Theta_i)$ and $\mu(\Theta_i)$ are positive for  $0<\Theta_i<1$. 

If one considers small perturbations to planar interface moving with a speed one, in the scaling induced by \eqref{eq:disp} the position of the interface is
\begin{equation}
\label{eq:ebn}
y=t+\delta\phi\left(\sqrt{\delta}x,\delta^2t\right),
\end{equation}
when viewed as a graph of a function in Cartesian coordinates.  The dispersion relation \eqref{eq:disp} strongly suggests that, at leading order, the function $\phi$ satisfies the classical equation  for
phase turbulence \cite{gr_rev,Kuramoto} given by
\begin{eqnarray}\label{eq:ks}
\phi_\tau-\frac12( \phi_\xi)^2+\lambda \phi_{\xi\xi}+\mu \phi_{\xi\xi\xi\xi}=0,
\end{eqnarray}
where $\xi=\sqrt{\delta}x$ and $\tau=\delta^2t$ represent scaled space and time variables. As a byproduct of analysis in this paper, we show that this is indeed the case.  \tb{The connection between the dispersion relation \eqref{eq:disp} and the equation \eqref{eq:ks} was first discussed in
\cite{gr_aa,gr_rev}.}

The spatio-temporal scaling in \eqref{eq:ebn} is natural for analysis of interface dynamics with infinitesimal deviation of the Lewis number from the stability threshold. However, in many practical situations, it is of interest to consider the behavior of an interface when a relative deviation of the Lewis number down from its critical value is small but finite. In this case the problem does not possess a convenient physical small parameter. Since it is often desirable to analyze combustion interfaces that are initially almost planar, the natural small parameter of a geometric origin is the  curvature of the interface. To make the appropriate scalings precise, let $0<\eps\ll 1$ and suppose that for any $0\leq t<T/\eps$ the following assumptions hold: (i) The curvature of the interface $\kappa=O(\eps)$; (ii) The normal velocity of the interface $V_n=O(1)$. In particular, these conditions are realized in the special case when the interface is a graph of a function in Cartesian coordinates, where the position function is given by
\begin{equation}
\label{eq:zop}
y=t+\frac{1}{\eps}\psi\left(\xi,\tau\right),\quad \xi=\eps x, \quad \tau=\eps t,
\end{equation}
and $\psi$ is a sufficiently smooth function that changes on a scale of order one. Note that in this case the slope of the interface is of order $O(1)$.

Given the scaling above, we find it most convenient to work with interface-attached coordinates.
We assume that the level set $\Theta=\Theta_i$ is a sufficiently smooth curve $\mathcal{C}$.
The interface is given by the position vector ${\bf R}(s,t),$ parametrized with respect to arc-length $s$ at a current time $t$.
The position of any point on a plane is then given by
\begin{eqnarray} 
{\bf r}(n,s,t)={\bf R}(s,t)+n{\bf N}(s,t).
\end{eqnarray}
Here, ${\bf N}$ is a normal to the interface and $n$ is a distance from the interface to the reference point.

In the new coordinates, the governing equations read:
\begin{eqnarray}
\label{whatever1}
&& \frac{\partial{\Theta_+}}{\partial{t}}=\Delta \Theta_+, \nonumber \\
&& \frac{\partial\Phi_+}{\partial t}=\rm{Le}^{-1} \Delta \Phi_+,
\end{eqnarray}
for $n>0$ and
\begin{eqnarray}
&&  \frac{\partial{\Theta_-}}{\partial{t}}=\Delta \Theta_-+A\Phi_-, \nonumber \\
&& \frac{\partial\Phi_-}{\partial t}=\rm{Le}^{-1} \Delta \Phi_--A\Phi_-,
\end{eqnarray}
for $n<0$.

This system of equation is complemented by the conditions far ahead 
\begin{eqnarray}
\Theta_+(n,s,t)\to 0, \qquad \Phi_+(n,s,t)\to 1 \quad \mbox{as} \quad n\to +\infty,
\end{eqnarray}
and far behind
\begin{eqnarray}
\Theta_-(n,s,t)\to 1, \qquad \Phi_-(n,s,t)\to 0 \quad \mbox{as} \quad n\to- \infty,
\end{eqnarray}
the interface, along with the continuity conditions on the interface
\begin{eqnarray}
\label{whatever2}
\lim_{n\to0} \Theta_+=\lim_{n\to0} \Theta_-=\Theta_i, \qquad \lim_{n\to 0} {\bf N}\cdot \nabla(\Theta_+-\Theta_-)=0.
\end{eqnarray}

In this paper we derive an asymptotic model for the dynamics of the interface associated with the scalings in \eqref{eq:zop} and establish a strongly nonlinear dependence of the interface velocity on curvature and its derivatives. It is important to note that this model is substantially more nonlinear than the one associated with the scaling in \eqref{eq:ebn}. In this regard, the model based on scaling \eqref{eq:ebn} can be viewed as weakly nonlinear, whereas the model based on scaling \eqref{eq:zop} is strongly nonlinear. We note that derivation of equation for dynamics of the diffusive interface under assumptions of this paper in the conventional thermo-diffusive model in high-activation-energy limit was performed in  \cite{FS87}. \tb{Further,
the nonlinear stability analysis of the model \eqref{eq:1}-\eqref{eq:3} with the zero-order reaction mechanism is presented in the recent paper \cite{brun}.}

Our purpose in deriving a strongly nonlinear reduced model is multifold. First, direct numerical simulations of cellular instabilities using the full system of equations \eqref{eq:1}-\eqref{eq:3} require substantial computational resources. As follows from linear stability analysis \cite{BGKS15}, perturbations of the interface induce extremely slowly decaying tails in the temperature and concentration fields. Thus even the simplest instability regime can be captured only in domains that are orders of magnitude larger than the thickness of the interface.  

Second, the model allows to relate the geometric and material characteristics of the problem with a dynamics of the interface.

Finally, the nonlinear model is derived in this paper under the weakest assumptions on scalings that allows for the dimensional reduction. Thus all other asymptotic model can be obtained from the present setup by introducing appropriate rescalings.

Note that  the derivation of our asymptotic model is free of an assumption that  the interface is given by a graph of a function.
This is rather useful and important feature of this model.
Indeed, in certain parameter regimes, when an interface initially is a graph of a function, it fails this property at some point during the evolution.


The paper is organized as follows. In Section 2, we present a formal asymptotic procedure that is subsequently used in Section 3 to obtain an asymptotic expression for the normal velocity of the combustion interface. In Section 4, we discuss the properties of the reduced model and present its further simplifications.
In Section 5 we present the results of numerical simulations of the reduced model. 
In Section 6 we summarize main results obtained in this paper.
In two appendices, we review some standard facts from differential geometry needed for derivation of the asymptotic model and provide the details of the computational setup.

\section{Scalings and asymptotic procedure}

In this section, we outline the asymptotic procedure that gives approximate solutions of the system \eqref{whatever1}-\eqref{whatever2} for the scalings discussed in the preceding section. In particular, we set
\begin{equation}
\begin{array}{rcl}
\tau&=&\eps t,\\
\sigma&=&\eps s,\\
\kappa&=&\eps\mathcal{K},
\end{array}
\end{equation}
where $\kappa$ is the curvature of $\mathcal{C}$ in the unscaled coordinates.  We now seek an asymptotic solution of \eqref{whatever1}-\eqref{whatever2} in the following form 
\begin{align}
\label{vn}
& \Theta_\pm=\Theta_{\pm}^{(0)}+\eps\Theta_{\pm}^{(1)}+\eps^2\Theta_{\pm}^{(2)}+ \eps^3\Theta_{\pm}^{(3)}+\mathcal{O}(\eps^4), \nonumber \\
& \Phi_{\pm}=\Phi_{\pm}^{(0)} + \eps \Phi_{\pm}^{(1)} + \eps^2 \Phi_{\pm}^{(2)}+\eps^3 \Phi_{\pm}^{(3)} + \mathcal{O}(\eps^4), \\
& V_n=1 + \eps V_n^{(1)} + \eps^2 V_n^{(2)} + \eps^3 V_n^{(3)}+\mathcal{O}(\eps^4),\nonumber 
\end{align}
where $\Theta_{\pm}^{(0)}$ and $\Phi_{\pm}^{(0)}$ correspond to planar interface solutions of \eqref{whatever1}-\eqref{whatever2} moving with velocity $1$. We also note that the number of terms in the expansions is chosen so as to guarantee well-posedness of the linearized version of the resulting asymptotic model.  \tb{Namely, the formal linearization of the resulting asymptotic model around the trivial solution produces instability only in a bounded range of
wave numbers.}

Substituting the expressions \eqref{vn} into \eqref{whatever1}-\eqref{whatever2}, collecting terms corresponding to different powers of $\eps$, and using the standard identities of differential geometry (cf. Appendix A), we obtain a recurrent system of ordinary differential equations for the temperature $\Theta_\pm$ and concentration  $\Phi_\pm$ fields, respectively, for $n>0$ (ahead of the interface)
\begin{equation}
\begin{cases}
\label{chuvak1}
\dfrac{\partial^2}{\partial n^2}\Theta_+^{(k)} +\dfrac{\partial }{\partial n}\Theta_+^{(k)}=f_+^{(k)}, \\\\ \dfrac{1}{\mathrm{Le}}\dfrac{\partial^2}{\partial n^2}\Phi_+^{(k)} +\dfrac{\partial }{\partial n}\Phi_+^{(k)}=g_+^{(k)},
\end{cases}
\end{equation}
and for $n<0$ (behind the interface)
\begin{equation}
\begin{cases}
\label{chuvak2}
\dfrac{\partial^2}{\partial n^2}\Theta_-^{(k)} +\dfrac{\partial }{\partial n}\Theta_-^{(k)}+A\Phi_-^{(k)}=  f_-^{(k)}, \\\\ \dfrac{1}{\mathrm{Le}}\dfrac{\partial^2}{\partial n^2}\Phi_-^{(k)}+\dfrac{\partial }{\partial n}\Phi_-^{(k)}-A\Phi_-^{(k)}= g_-^{(k)},
\end{cases}
\end{equation}
Here $k=0,\ldots,3$ and the right hand sides of the equations in \eqref{chuvak1}-\eqref{chuvak2} are given by the following expressions
\begin{equation}
f_\pm^{(0)}=0,\qquad  g_\pm^{(0)}=0,
\end{equation}
at $\mathcal O(1)$,
\begin{equation}
f_\pm^{(1)}= -\left(V_n^{(1)}+{\mathcal{K}}\right)\dfrac{\partial\Theta_\pm^{(0)}}{\partial n}, \qquad g_\pm^{(1)}=-\left(V_n^{(1)}+\dfrac{{\mathcal{K}}}{\mathrm{Le}}\right)\dfrac{\partial \Phi_\pm^{(0)}}{\partial n},
\end{equation}
at $\mathcal O(\eps)$,
\begin{eqnarray}
& f_\pm^{(2)}= -\left(V_n^{(1)}+{\mathcal{K}}\right)\dfrac{\partial\Theta_\pm^{(1)}}{\partial n}+\left(n{\mathcal{K}}^2-V_n^{(2)}\right)\dfrac{\partial\Theta_\pm^{(0)}}{\partial n}+\mathcal{L}_\tau\Theta_\pm^{(1)},& \\
& g_\pm^{(2)}= -\left(V_n^{(1)}+\dfrac{{\mathcal{K}}}{\mathrm{Le}}\right)\dfrac{\partial \Phi_\pm^{(1)}}{\partial n}+\left(\dfrac{n{\mathcal{K}}^2}{\mathrm{Le}}-V_n^{(2)}\right)\dfrac{\partial \Phi_\pm^{(0)}}{\partial n}+\mathcal{L}_\tau \Phi_\pm^{(1)}, &
\end{eqnarray}
at $\mathcal O(\eps^2)$,
\begin{eqnarray}
&f_\pm^{(3)}= -\left(V_n^{(1)} + {\mathcal{K}}\right)\dfrac{\partial\Theta_\pm^{(2)}}{\partial n} + \left(n{\mathcal{K}}^2-V_n^{(2)}\right)\dfrac{\partial\Theta_\pm^{(1)}}{\partial n} - \left(V_n^{(3)} + n^2{\mathcal{K}}^3\right)\dfrac{\partial\Theta^{(0)}_\pm}{\partial n} + \mathcal{L}_\tau\Theta_\pm^{(2)}-\dfrac{\partial^2\Theta^{(1)}_\pm}{\partial \sigma^2}, & \\
&g_\pm^{(3)}=-\left(V_n^{(1)}+\dfrac{{\mathcal{K}}}{\mathrm{Le}}\right)\dfrac{\partial \Phi_\pm^{(2)}}{\partial n} + \left(\dfrac{n{\mathcal{K}}^2}{\mathrm{Le}}-V_n^{(2)}\right)\dfrac{\partial \Phi_\pm^{(1)}}{\partial n} -\left(V_n^{(3)} + \dfrac{n^2{\mathcal{K}}^3}{\mathrm{Le}}\right)\dfrac{\partial \Phi^{(0)}_\pm}{\partial n} + \mathcal{L}_\tau \Phi_\pm^{(2)} - \dfrac{1}{\mathrm{Le}}\dfrac{\partial^2 \Phi^{(1)}_\pm}{\partial \sigma^2},&
\end{eqnarray}
at $\mathcal O(\eps^3)$.

The governing equations are supplemented by the far-field conditions
\begin{eqnarray}
&\displaystyle\lim\limits_{n \to \infty}\Theta_+^{(0)} = 0, \qquad \qquad\displaystyle\lim\limits_{n \to -\infty}\Theta_-^{(0)} = 1,& \label{1bc} \\
&\displaystyle\lim\limits_{n \to \infty}\Phi_+^{(0)} = 1, \qquad \qquad\displaystyle\lim\limits_{n \to -\infty}\Phi_-^{(0)} = 0,&\label{1bca}
\end{eqnarray}
at $\mathcal{O}(1)$ and
\begin{eqnarray}
&\displaystyle\lim\limits_{n \to \infty}\Theta_+^{(k)} = 0,\qquad \Theta_-^{(k)}\mbox{ grows not faster than a polynomial as }n\to-\infty, \label{hbc1} \\
&\displaystyle\lim\limits_{n \to \infty}\Phi_+^{(k)} = 0, \qquad \qquad \displaystyle\lim\limits_{n \to -\infty}\Phi_-^{(k)}=0,&\label{1bca}
\end{eqnarray}
at higher orders.

In addition, we also impose continuity conditions on the interface $\mathcal C$ on the solution and its normal derivates 
\begin{equation}
\left[\Theta^{(k)}\right]_{n=0}=0,\qquad \left[\Phi^{(k)}\right]_{n=0}=0,\qquad \left[\frac{\partial\Theta^{(k)}}{\partial n}\right]_{n=0}=0,\qquad \left[\frac{\partial\Phi^{(k)}}{\partial n}\right]_{n=0}=0,
\end{equation}
where $\left[\cdot\right]_{n=0}$ stands for a jump of a quantity when crossing the interface.

Note that the interface $\mathcal C$ corresponds to the level set $\Theta=\Theta_i$ and, therefore
\begin{equation}
\Theta_\pm^{(0)}|_{n=0}=\Theta_i,\qquad \Theta_\pm^{(k)}|_{n=0}=0\ \mbox{for}\ k=1,2,3.
\end{equation}

In the next section we derive an asymptotic expression for the normal velocity of the interface.

\section{Normal velocity of the interface}

\tb{In this section we present the principal terms in the asymptotic expansion of the normal velocity, temperature, and concentration that were obtained by following the steps outlined in the previous section. These terms were  obtained  and verified using symbolic computations in Wolfram Mathematica.}

We start by observing that planar traveling interface solution of the system of governing equations is given by 
\begin{equation}
     \begin{aligned}
      \Theta^{(0)}_+&= \Theta_i\exp(-n), \\
      \Theta^{(0)}_-&=1+(\Theta_i-1)\exp\left(\dfrac{\Theta_i}{1-\Theta_i}n\right),
   \end{aligned}
  \label{W}
\end{equation}
for the temperature and by
\begin{equation}
     \begin{aligned}
      \Phi^{(0)}_+&= 1-\dfrac{\Theta_i}{\mathrm{Le}+\Theta_i-\mathrm{Le}\,\Theta_i}\exp(-\mathrm{Le}\, n), \\
      \Phi^{(0)}_-&=\dfrac{\mathrm{Le}(1-\Theta_i)}{\mathrm{Le}+\Theta_i-\mathrm{Le}\,\Theta_i}\exp\left(\dfrac{\Theta_i}{1-\Theta_i}n\right),
   \end{aligned}
  \label{W}
\end{equation}
for the concentration.
Note that these equations are identical to \eqref{eq:4}-\eqref{eq:5}.

At order $\mathcal{O}(\eps)$ temperature and concentration are given by 
\begin{equation}
\begin{aligned}
\Theta_+^{(1)}&={\mathcal{K}}\dfrac{n{\Theta_i}^2(\mathrm{Le}-1)(\mathrm{Le}(\Theta_i-1)-2)}{2\mathrm{Le}(\Theta_i-1)(\mathrm{Le}+\Theta_i-\mathrm{Le}\Theta_i)}\exp(-n), \\
\Theta_-^{(1)}&={\mathcal{K}}(\mathrm{Le}-1)\left(\dfrac{2(\mathrm{Le}+\Theta_i-\mathrm{Le}\Theta_i)}{2\mathrm{Le}(\mathrm{Le}(\Theta_i-1)-\Theta_i)}-\dfrac{2\Theta_i+\mathrm{Le}(2+\Theta_i(n(\Theta_i-2)-2)))}{2\mathrm{Le}(\mathrm{Le}(\Theta_i-1)-\Theta_i)}\exp\left(\dfrac{\Theta_i n}{1-\Theta_i}\right)\right),
\end{aligned}
\end{equation}
and 
\begin{equation}
\begin{aligned}
\Phi_+^{(1)}&={\mathcal{K}}\dfrac{\Theta_i(\mathrm{Le}-1)(\Theta_i-2)(\mathrm{Le}(\Theta_i-1)-2-2\Theta_i(n-1))}{2(\Theta_i-1)^2(\mathrm{Le}+\Theta_i-\mathrm{Le}\Theta_i)^2}\exp(-\mathrm{Le}~n), \\
\Phi_-^{(1)}&={\mathcal{K}}\Theta_i\dfrac{(\mathrm{Le}-1)(\Theta_i-2)(\mathrm{Le}~ n-2)}{2(\mathrm{Le}+\Theta_i-\mathrm{Le}~\Theta_i)^2}\exp\left(\dfrac{\Theta_i n}{1-\Theta_i}\right),
\end{aligned}
\end{equation}
respectively. The first correction to the normal velocity then reads
\begin{equation}
V_n^{(1)}=\gamma_1{\mathcal{K}}\label{v1},
\end{equation}
where
\begin{equation}
\gamma_1(\mathrm{Le}, \Theta_i):=\frac{1}{2}\left(\dfrac{2-2\mathrm{Le}}{\mathrm{Le}-\mathrm{Le}~\Theta_i}-\dfrac{2-\mathrm{Le}}{\mathrm{Le}+\Theta_i-\mathrm{Le}~\Theta_i}-1\right).
\end{equation}
One can easily verify that $\gamma_1(\mathrm{Le_0(\Theta_i)}, \Theta_i)=0$ and the sign of $\gamma_1(\mathrm{Le}, \Theta_i)$ is the same as the sign of $\Le-\Le_0$, where $\Le_0$ is a critical Lewis number defined in \eqref{le0}. This observation is precisely what can be expected from the linear stability analysis. Furthermore, when $1>\Le>\Le_0$ the asymptotic behavior of the normal velocity can be adequately described by the following equation 
\begin{equation}
\label{eq:mdl}
V_n=1+\eps\gamma_1\mathcal{K}.
\end{equation}
Note that \eqref{eq:mdl} is linearly well-posed and requires no additional regularization. However, for $\Le\leq\Le_0$, the equation \eqref{eq:mdl} is linearly ill-posed and must be regularized by retaining more terms in a relevant asymptotic expansion. These terms in the expansion of the normal velocity are given by 
\begin{equation}
\label{eq:v2}
V_n^{(2)}=\gamma_2{\mathcal K}^2,
\end{equation}
at $\mathcal{O}(\eps^2)$ and 
\begin{equation}
\label{eq:v3}
V_n^{(3)}=\gamma_3{\mathcal K}^3 + \gamma_4\dfrac{\partial^2{\mathcal K}}{\partial \sigma^2},
\end{equation}
at $\mathcal{O}(\eps^3)$.
Here the coefficients $\gamma_2$, $\gamma_3$, and $\gamma_4$ are given by
\begin{equation}
\begin{aligned}
\gamma_2(\mathrm{Le}, \Theta_i):=&\left\{8 \mathrm{Le} (\Theta_i-1) (\mathrm{Le} (\Theta_i-1)-2 \Theta_i) (\mathrm{Le}-\mathrm{Le}\Theta_i+\Theta_i)^2\right\}^{-1}\\&\left\{\Theta_i (\mathrm{Le}^4 (\Theta_i-1)^2 (3 \Theta_i-4)-2 \mathrm{Le}^3 \left(6 \Theta_i^3-13 \Theta_i^2+5 \Theta_i+2\right)\right.\\&\left.+\mathrm{Le}^2 \left(15 \Theta_i^3-14 \Theta_i^2-21 \Theta_i+12\right)+\mathrm{Le} \left(-6 \Theta_i^3+6 \Theta_i^2+4 \Theta_i+12\right)-8 \Theta_i^2)\right\},
\end{aligned}
\end{equation}
\begin{equation}
\begin{aligned}
\gamma_3(\mathrm{Le}, \Theta_i):=-&\left\{8 \mathrm{Le}^3 (\Theta_i-1)^3 (\mathrm{Le} - \mathrm{Le}\Theta_i + \Theta_i)^3\right\}^{-1} \\ & \left\{(\mathrm{Le}-1)^2 \Theta_i (\mathrm{Le}^4 \left(1-\Theta_i\right)^3 ((\Theta_i-6) \Theta_i+4)+\mathrm{Le}^3 (\Theta_i-1)^2 (\Theta_i+2) ((\Theta_i-9) \Theta_i+6)\right. \\ &\left.+2 \mathrm{Le}^2 (\Theta_i-1) (\Theta_i (\Theta_i (3 \Theta_i-1)+16)-12)\right.\\ &\left.+4 \mathrm{Le} \Theta_i (\Theta_i ((13-3 \Theta_i) \Theta_i-24)+12)+8 (\Theta_i-2)^2 \Theta_i^2)\right\},
\end{aligned}
\end{equation}
and
\begin{equation}
\begin{aligned}
\gamma_4(\mathrm{Le}, \Theta_i):=-&\left\{8 \mathrm{Le}^3 (\Theta_i-1)^3 (\mathrm{Le} (\Theta_i-1)-2 \Theta_i)^3 (\mathrm{Le} (\Theta_i-1)-\Theta_i)^3\right\}^{-1} \\ &\left\{\mathrm{Le}^9 (\Theta_i-1)^6 (3 \Theta_i^3-26 \Theta_i^2+40 \Theta_i-16)-\mathrm{Le}^8 (\Theta_i-1)^5 \Theta_i \left(27 \Theta_i^3-251 \Theta_i^2+388 \Theta_i-152\right)\right.\\&+\mathrm{Le}^7 (\Theta_i-1)^4 \Theta_i \left(99 \Theta_i^4-1012 \Theta_i^3+1587 \Theta_i^2-626 \Theta_i+16\right)\\&-\mathrm{Le}^6 (\Theta_i-1)^3 \Theta_i \left(189 \Theta_i^5-2219 \Theta_i^4+3551 \Theta_i^3-1321 \Theta_i^2-72 \Theta_i+56\right)\\&+2 \mathrm{Le}^5 (\Theta_i-1)^2 \Theta_i \left(99 \Theta_i^6-1478 \Theta_i^5+2584 \Theta_i^4-1150 \Theta_i^3+133 \Theta_i^2-80 \Theta_i+36\right)\\&-4 \mathrm{Le}^4 \Theta_i^2 \left(27 \Theta_i^7-674 \Theta_i^6+2091 \Theta_i^5-2554 \Theta_i^4+1675 \Theta_i^3-852 \Theta_i^2+351 \Theta_i-64\right)\\&+8 \mathrm{Le}^3 \Theta_i^3 \left(3 \Theta_i^6-205 \Theta_i^5+685 \Theta_i^4-891 \Theta_i^3+672 \Theta_i^2-320 \Theta_i+64\right)\\&+32 \mathrm{Le}^2 \Theta_i^4 \left(26 \Theta_i^4-103 \Theta_i^3+130 \Theta_i^2-80 \Theta_i+21\right)\\&\left.-32 \mathrm{Le} \Theta_i^5 \left(10 \Theta_i^3-43 \Theta_i^2+41 \Theta_i-14\right)+64 \Theta_i^6 \left(\Theta_i^2-4 \Theta_i+2\right)\right\}.
\end{aligned}
\end{equation}
Note that we do not give the explicit expressions for higher-order corrections for either temperature or concentration profiles as the asymptotic procedure leading to these, although simple, is rather tedious. 

\begin{figure}[!]
\centering
\includegraphics[height=3in]{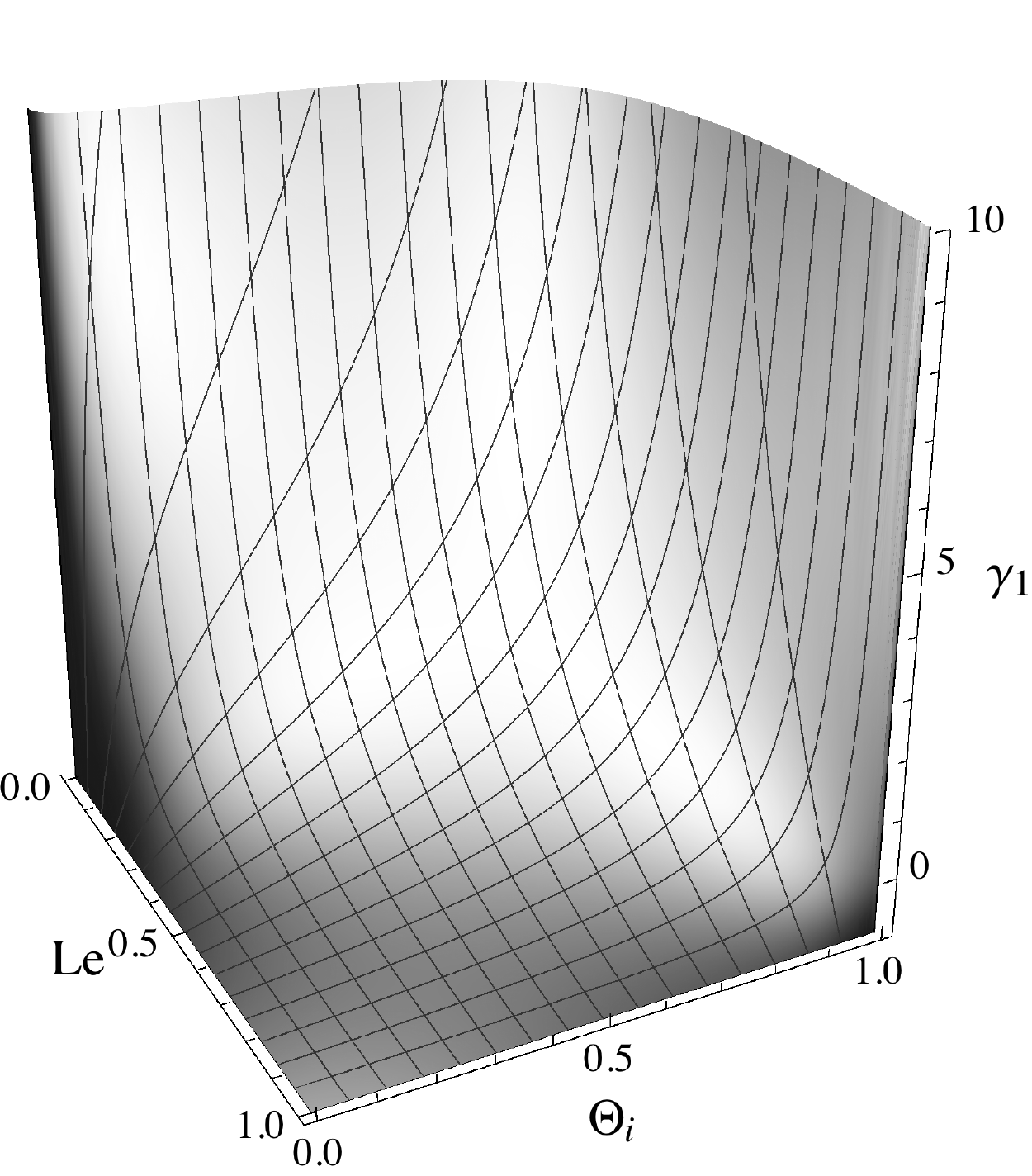}
\caption{Coefficient $\gamma_1$ as a function of the ignition temperature $\Theta_i$ and the Lewis number $\Lw$.}
  \label{fig:g1}
\end{figure}

\begin{figure}[!]
\centering
\includegraphics[height=3in]{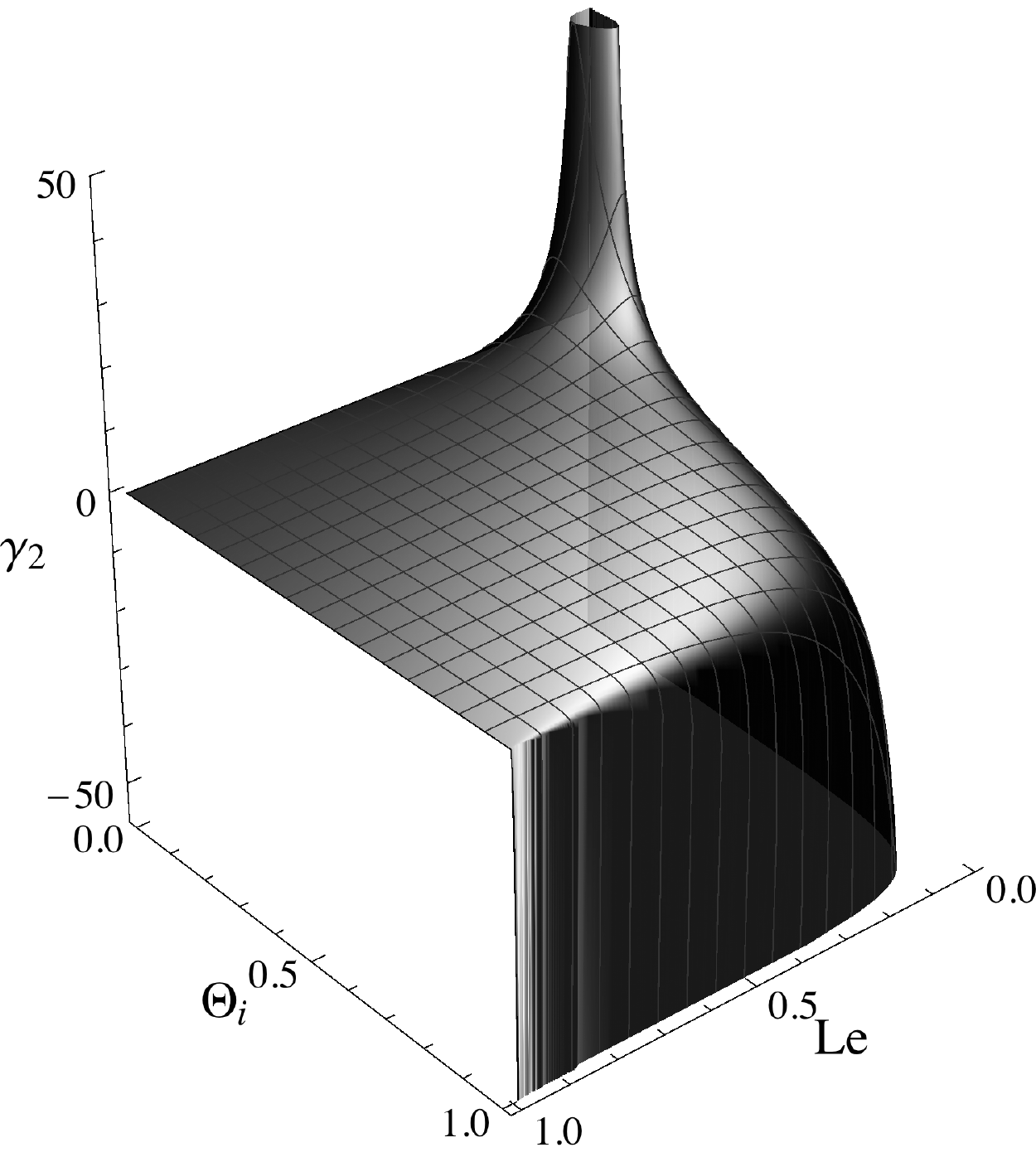}
\caption{Coefficient $\gamma_2$ as a function of the ignition temperature $\Theta_i$ and the Lewis number $\Lw$.}
  \label{fig:g2}
\end{figure}

\begin{figure}[!]
\centering
\includegraphics[height=3in]{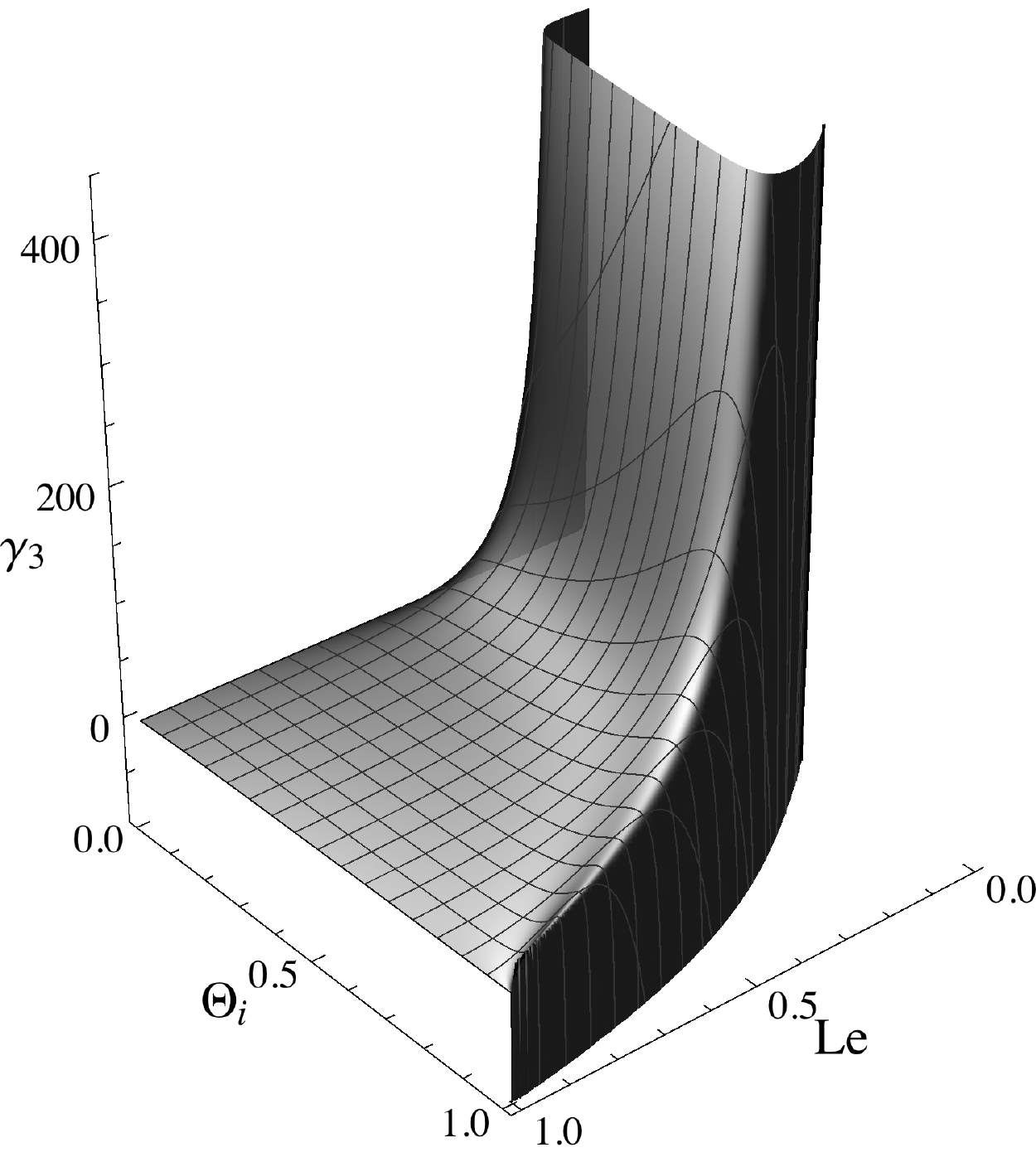}
\caption{Coefficient $\gamma_3$ as a function of the ignition temperature $\Theta_i$ and the Lewis number $\Lw$.}
  \label{fig:g3}
\end{figure}

\begin{figure}[!]
\centering
\includegraphics[height=3in]{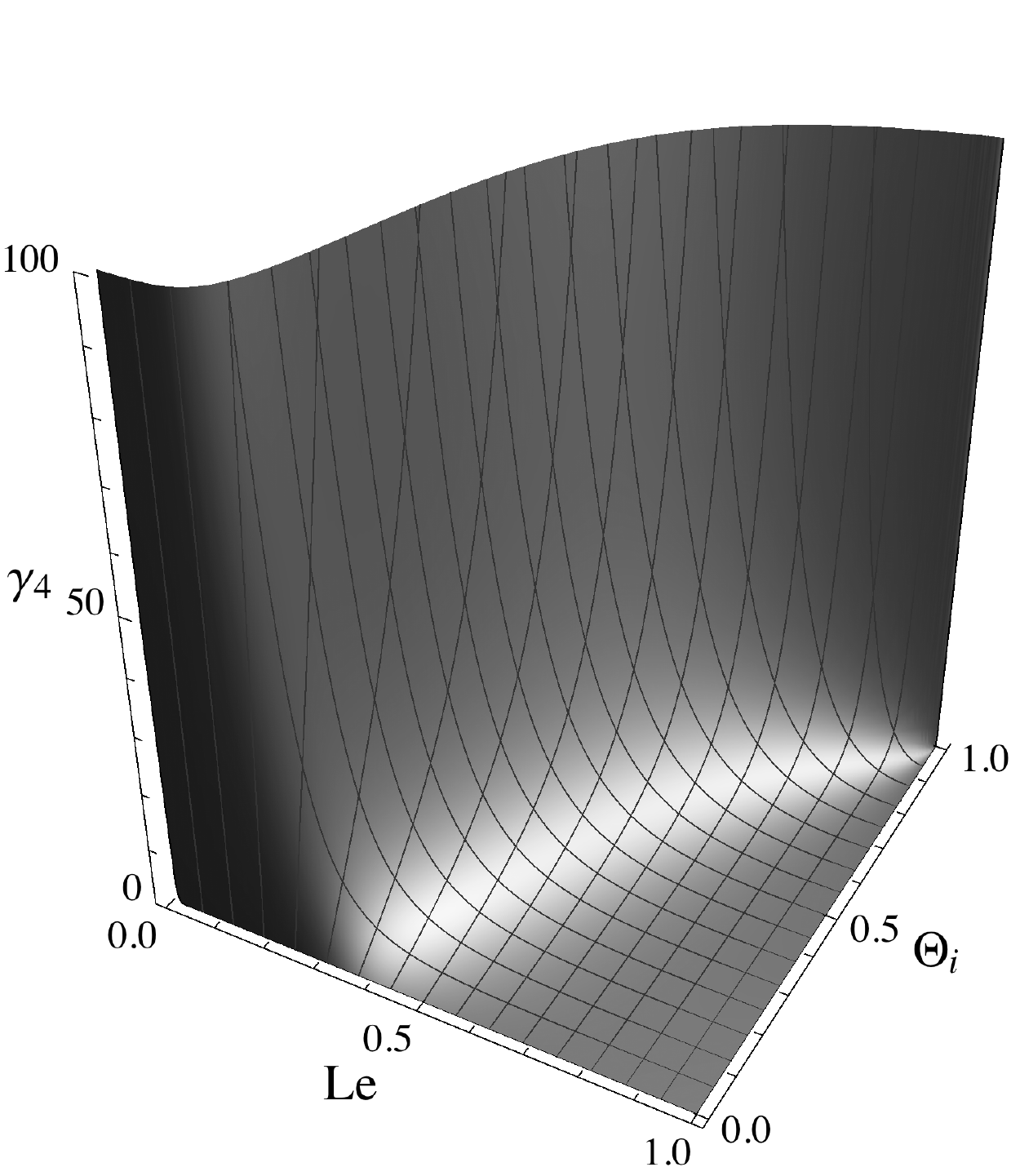}
\caption{Coefficient $\gamma_4$ as a function of the ignition temperature $\Theta_i$ and the Lewis number $\Lw$.}
  \label{fig:g4}
\end{figure}

The dependencies of $\gamma_i$, $i=1,\ldots,4$ on the ignition temperature and the Lewis number are nontrivial, as can be seen in Figs \ref{fig:g1}-\ref{fig:g4}.
The expression for the normal velocity up to $\mathcal{O}(\eps^3)$ now takes the form
\begin{equation}
\label{eq:velo}
V_n=1 + \eps\gamma_1{\mathcal K} + \eps^2\gamma_2{\mathcal K}^2 + \eps^3\left(\gamma_3{\mathcal K}^3 + \gamma_4\dfrac{\partial^2{\mathcal K}}{\partial \sigma^2}\right).
\end{equation}

The following two observations are now in order. First, the terms  quadratic  and cubic in curvature have no influence on linear well-posedness of the problem. Second, the coefficient $\gamma_4$ is strictly positive for all relevant parameter values, which ensures linear well-posedness of \eqref{eq:velo}.
\tb{We also note that the level set $\gamma_1=0$ that separates linearly stable and linearly unstable regions  in the parametric space is depicted in Fig. \ref{fig:st}.}

\section{Properties of the asymptotic model and its further simplifications}

In this section we explore the properties of the asymptotic model \eqref{eq:velo}, consider further simplifications of \eqref{eq:velo}, and establish the relationship between \eqref{eq:velo} and the model \eqref{eq:ks}. 

Recall that the asymptotic reduction of \eqref{eq:1}-\eqref{eq:3} to the equation \eqref{eq:velo} of motion of the combustion interface was obtained in this work under the single assumption of the small interface curvature. 
The derivation did not impose any restrictions on values of the parameters of the problem, such as the ignition temperature and the Lewis number. 

Since the smallness of the curvature is an ingredient, needed to derive any asymptotic interface model that effectively replaces \eqref{eq:1}-\eqref{eq:3}, the equation \eqref{eq:velo} is the most general in this class of models. Consequently, any other interface model can be obtained from \eqref{eq:velo} by introducing an appropriate rescaling.
We note that this asymptotic model  is valid as long as the coefficients $\gamma_i$ are not too large, that is, their product with a typical scale of curvature variations
remains small. This is always the case during some initial stage of evolution for almost planar interfaces. Therefore, the model \eqref{eq:velo} provides most general asymptotic description of a combustion interface at least for some (possibly short) time interval. For longer times, the predictions of  \eqref{eq:velo} are not necessarily accurate, even if a solution of this problem exists globally in time.

Let us now discuss  simplifications of model \eqref{eq:velo} corresponding to certain parameter regimes. We will consider a situation when the interface position
is given by a graph of the function $\psi(\xi,\tau)$ defined in \eqref{eq:zop}. Since the model \eqref{eq:velo} is  based on the scaling introduced in \eqref{eq:zop}, we obtain the following expressions in terms of the function $\psi$ for the quantities involved in \eqref{eq:velo}:
\begin{eqnarray}\label{eq:k1}
V_n=\frac{1+ \psi_\tau}{\sqrt{1+\psi_\xi^2}},
\end{eqnarray}
\begin{eqnarray}\label{eq:k2}
{\mathcal K}=-\frac{\psi_{\xi\xi}}{(1+\psi_\xi^2)^{3/2}},
\end{eqnarray}
\begin{eqnarray}\label{eq:k3}
\dfrac{\partial^2{\mathcal K}}{\partial \sigma^2}=\frac{1}{\sqrt{1+ \psi_\xi^2}}\frac{\partial}{\partial \xi}\left(\frac{1}{\sqrt{1+ \psi_\xi^2}}\frac{\partial}{\partial \xi} {\mathcal K}\right).
\end{eqnarray}
We will consider two parameter regimes.

\medskip

{\bf Regime I}. First, consider a situation when the Lewis number slightly deviates from its critical value. To this end, we assume that  $\eps=\sqrt{\delta}$, where the positive $\delta\ll1$ is as in \eqref{eq:delta} and set 
\begin{eqnarray}
\psi(\xi,\tau)=\delta^{3/2} \chi(\xi, \bar \tau), \quad \bar \tau=\delta^{3/2} \tau.
\end{eqnarray}
In this scaling, we have
\begin{eqnarray}\label{eq:k1a}
V_n  \approx 1+\delta^3\left( \chi_{\bar \tau}-\frac12 \left(\chi_{\xi}\right)^2\right),
\qquad
{\mathcal K} \approx -\delta^{3/2} \chi_{\xi\xi},
\qquad
\dfrac{\partial^2{\mathcal K}}{\partial \sigma^2} \approx -\delta^{3/2} \chi_{\xi\xi\xi\xi}.
\end{eqnarray}
Substituting \eqref{eq:k1a} into \eqref{eq:velo} we obtain
\begin{eqnarray}\label{eq:106}
\chi_{\bar \tau}-\frac12 (\chi_{\xi})^2 =-\frac{\gamma_1}{\delta} \chi_{\xi\xi}-\gamma_4 \chi_{\xi\xi\xi\xi},
\end{eqnarray}
at leading order in $\delta$. Note that in the limit $\delta\to 0$ we have $\gamma_1/\delta \to \lambda$, $\gamma_4\to \mu$, where $\lambda$ and $\mu$ are given by \eqref{eq:lam} and \eqref{eq:mu}, respectively.  Therefore, in this limit we formally recover \eqref{eq:ks}. 

{\bf Regime II.} The second simplification of \eqref{eq:velo} employs the fact that 
$\gamma_1/\gamma_4$ is always small. \tb{Indeed as follows from straightforward computations, $|\gamma_1/\gamma_4|\le 0.22$
for all relevant values of $\Theta_i$ and $\Lw$. }
To take advantage of this, we set $\eps=\sqrt{\gamma_1/\gamma_4}$ and choose
\begin{eqnarray}
\psi(\xi,\tau)=\gamma_1 \eps h(\xi,\tilde \tau), \quad \tilde \tau =\gamma_1 \eps \tau.
\end{eqnarray}
As long as  $\gamma_1 \eps$ is sufficiently small (for example, in a vicinity of the stability threshold), we have
\begin{eqnarray}
V_n\approx 1+(\gamma_1\eps)^2 \left[h_{\tilde \tau}-\frac12 (h_{\xi})^2\right], \qquad {\mathcal K}\approx -\gamma_1 \eps h_{\xi\xi}, \qquad 
\dfrac{\partial^2{\mathcal K}}{\partial \sigma^2}\approx -\gamma_1 \eps h_{\xi\xi\xi\xi}.
\end{eqnarray}
Substituting these expressions into \eqref{eq:velo} we obtain
\begin{eqnarray}
h_{\tilde \tau}-\frac12 (h_{\xi})^2=-h_{\xi\xi}-h_{\xi\xi\xi\xi}+ p h_{\xi\xi}^2-q h_{\xi\xi}^3,
\end{eqnarray}
at leading order, where
\begin{eqnarray}
p:=\gamma_1 \gamma_2/\gamma_4, \quad q:=\gamma_1^3\gamma_3/\gamma_4^2,
\end{eqnarray}
are functions of $\Le$ and $\Theta_i$.
\begin{figure}[!]
\centering
\includegraphics[height=2in]{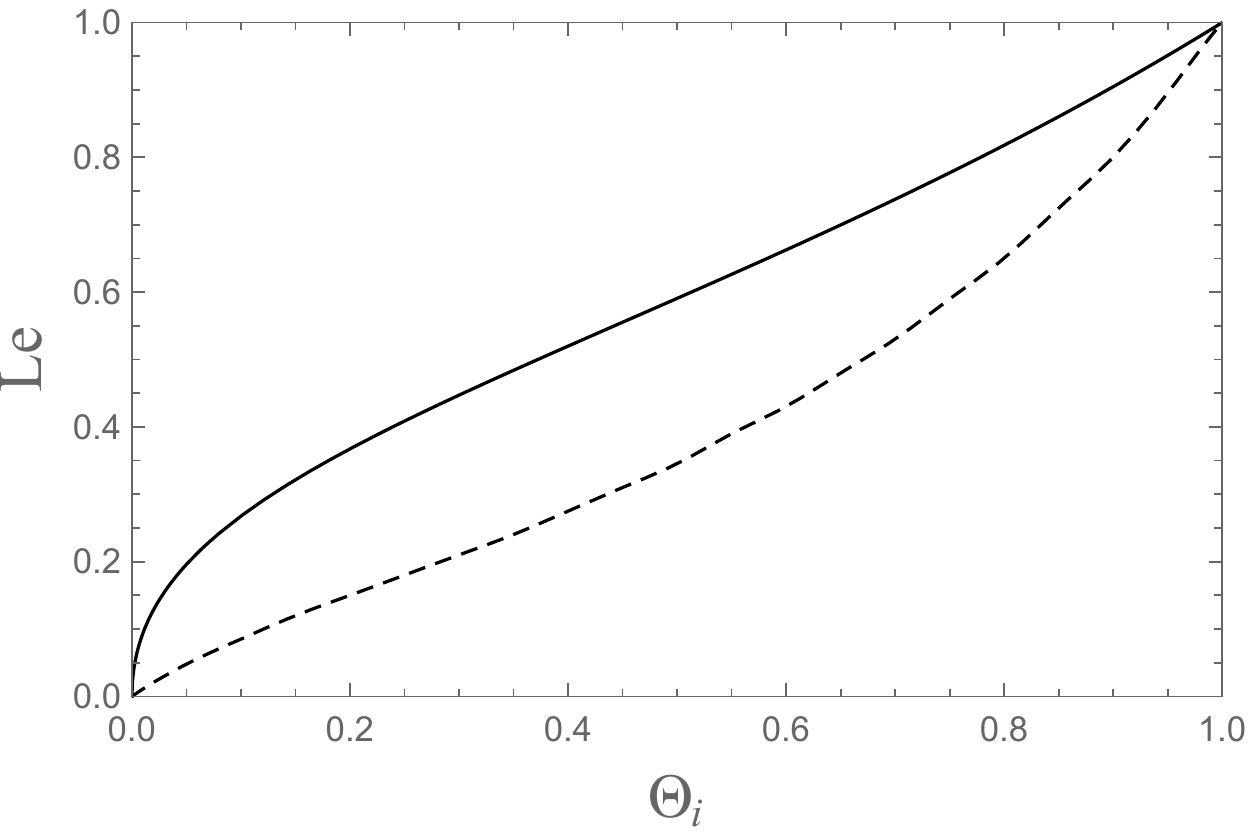}
\caption{Range of applicability of model \eqref{eq:ksa} with restrictions outlined in the text. The solid line represent critical value of the Lewis number. The dashed line
represent the boundary where the restrictions on the coefficients $p$ and $q$ become invalid. The range of applicability of \eqref{eq:ksa} is represented by a set bounded by the solid and dashed curves.}
  \label{fig:ksa}
\end{figure}
Straightforward computations show that both $|p|,|q|\le 1$ for all sub-critical values of the Lewis number and ignition temperature.
Moreover, one can verify that in a vicinity of the critical Lewis number both $p,q$ are negligibly small. Fig. \ref{fig:ksa} shows a portion of the parameter space
where $|p|,|q|\le 0.05$ and $\gamma_1 \eps \le 0.1$. In this part of the parameter space the equation \eqref{eq:velo} reduces to 
\begin{eqnarray}\label{eq:ksa0}
\tilde h_{\tilde \tau}-\frac12 (\tilde h_{\xi})^2=-\tilde h_{\xi\xi}-\tilde h_{\xi\xi\xi\xi}.
\end{eqnarray}
We conclude that in the immediate vicinity of the stability threshold, terms that depend on higher power of the curvature in the evolution equation 
 \eqref{eq:velo} can be neglected. Therefore, at least heuristically, in this region  the model \eqref{eq:velo} takes a particularly simple form
 \begin{equation}
\label{eq:ksa}
V_n=1 + \eps\gamma_1{\mathcal K}  + \eps^3 \gamma_4\dfrac{\partial^2{\mathcal K}}{\partial \sigma^2}.
\end{equation}
 
 In the next section we will compare the results of numerical simulations for the models \eqref{eq:106}, \eqref{eq:velo}, and \eqref{eq:ksa}.
 
 \section{Numerical results}
 \tb{In this section we use numerical simulations to examine the relationship between the solutions of the model \eqref{eq:velo} and its simplifications, given by \eqref{eq:106} and \eqref{eq:ksa}, respectively.  Note that we do not aim to present a comprehensive numerical investigation of possible parameter regimes, 
 but rather to provide guidance for further numerical and analytical explorations of \eqref{eq:velo}.}

 
 \subsection{Comparison between solutions of the models \eqref{eq:velo} and  \eqref{eq:106}} 
  In order to compare the predictions of the  model  \eqref{eq:velo} derived in this paper with these of its partially linearized version \eqref{eq:106}---identical in its structure to a well known equation for phase turbulence---we performed several numerical experiments using COMSOL \cite{Comsol}.  \tb{Representing the position of the interface as a graph of function $\psi$ (cf. \eqref{eq:zop}), we obtain a fourth order, nonlinear parabolic PDE for $\psi$ by substituting \eqref{eq:k1}--\eqref{eq:k3} into  \eqref{eq:velo}.   The resulting equation is then indistingushable from \eqref{eq:velo}, as long as the interface remains graph of a function.} The problem is then solved on a 
strip-like domain, subject to periodic boundary conditions.

\begin{figure}[H]
\centering
\includegraphics[height=2.in]{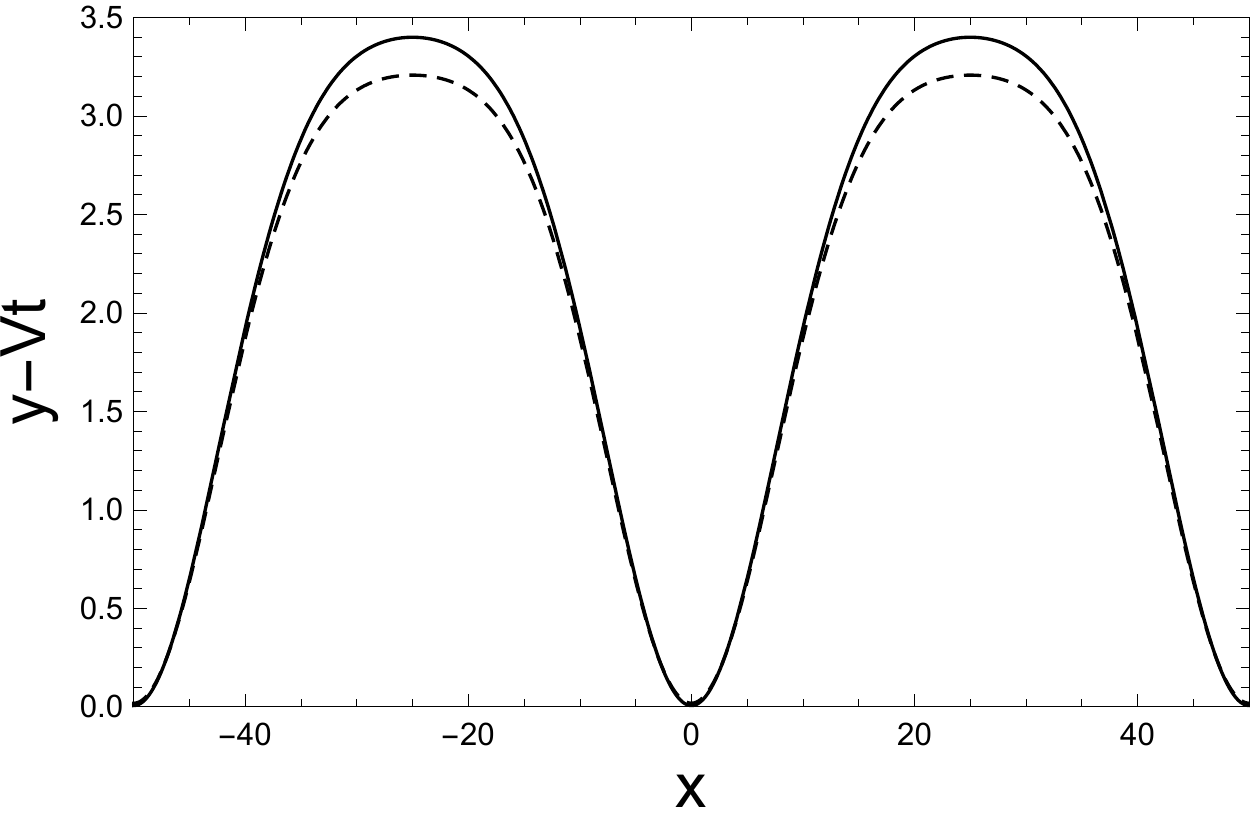}
\caption{Steady-state shape of the interface as predicted by \eqref{eq:106} (solid line) and \eqref{eq:velo} (dashed line) when $\Lw=0.73$ and $\Theta_i=0.8$. The parameter $\delta=0.11$. The  average flame speeds are $V=1.013$ and $V=1.012$, respectively.}
  \label{fig:1}
\end{figure}

\begin{figure}[H]
\centering
\includegraphics[height=2.in]{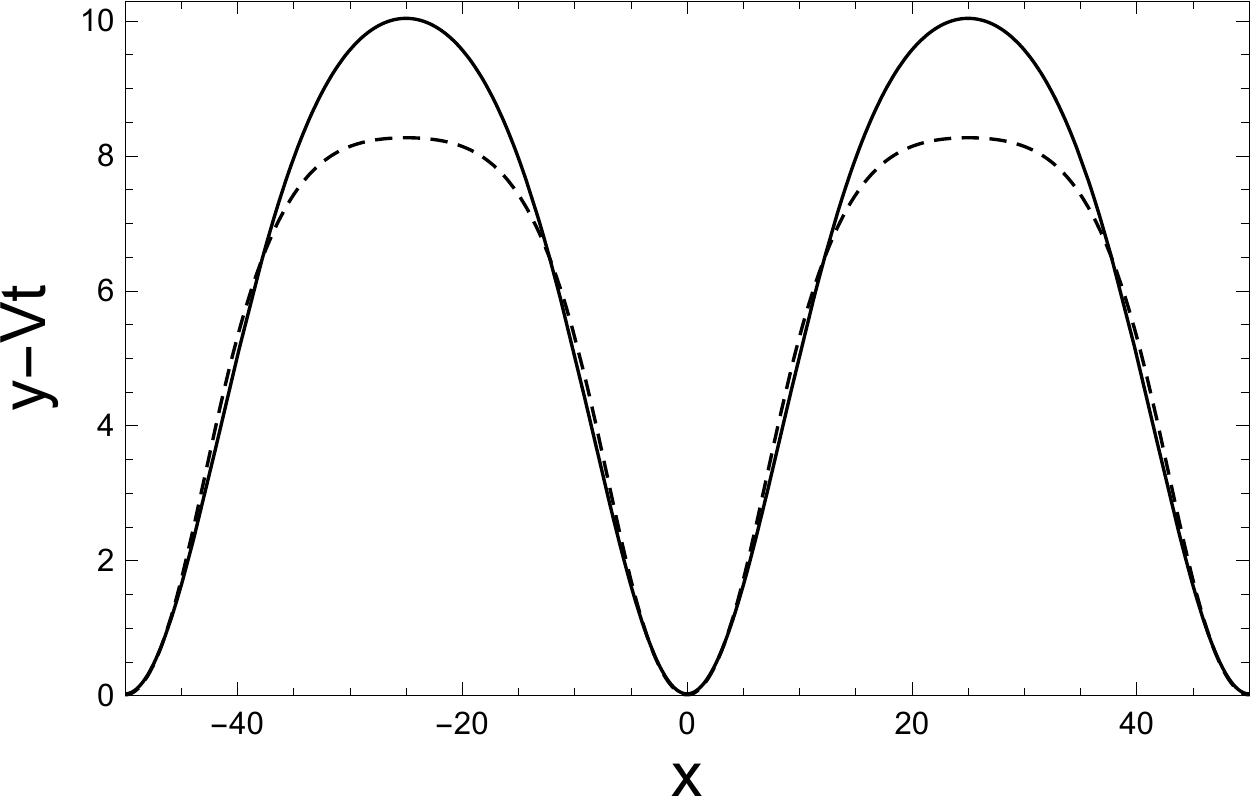}
\caption{Steady-state shape of the interface as predicted by \eqref{eq:106} (solid line) and \eqref{eq:velo} (dashed line) when $\Lw=0.63$ and $\Theta_i=0.8$. The parameter $\delta=0.23$. The  average flame speeds are $V=1.107$ and $V=1.077$, respectively.}
  \label{fig:2}
\end{figure}

\begin{figure}[H]
\centering
\includegraphics[height=2.in]{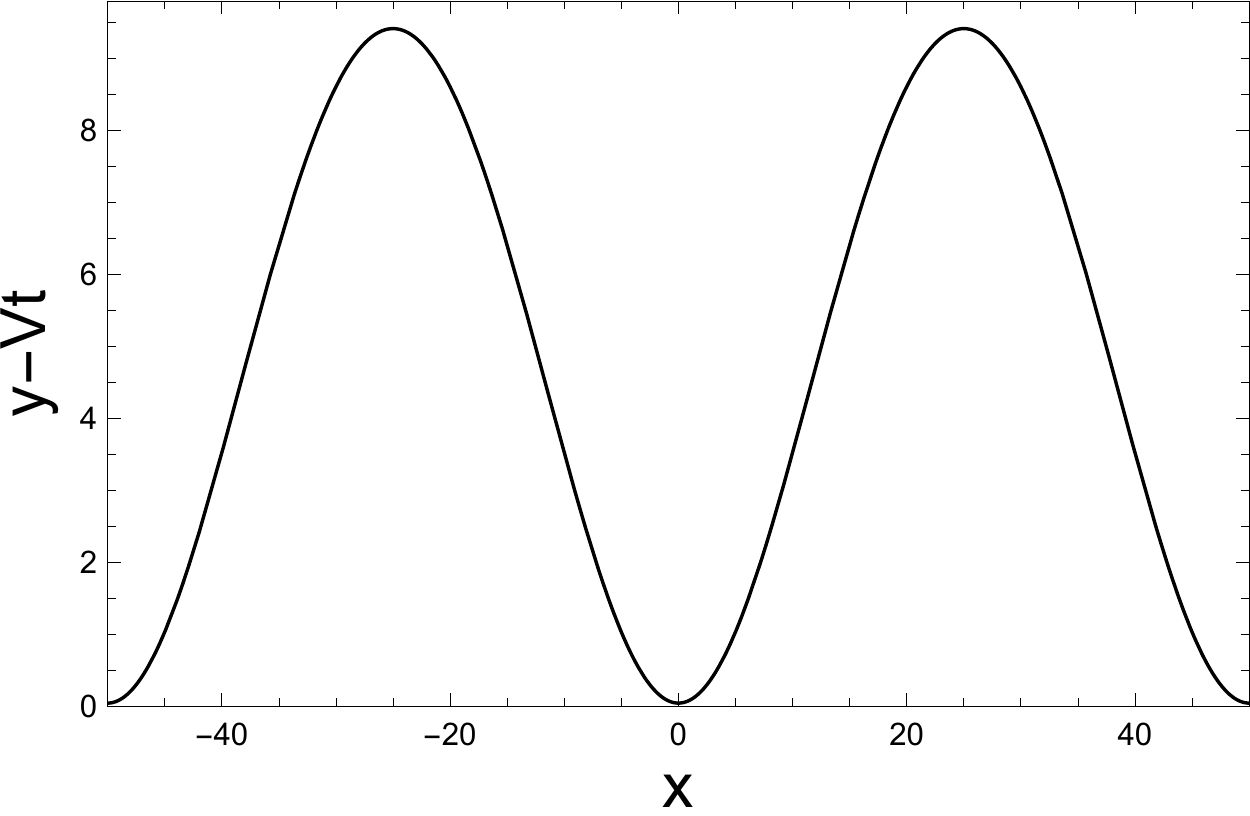}
\caption{Steady-state shape of the interface as predicted by \eqref{eq:106} when $\Lw=0.5$ and $\Theta_i=0.8$. The parameter $\delta=0.39$.}
  \label{fig:3}
\end{figure}

\begin{figure}[H]
\centering
\includegraphics[height=2.in]{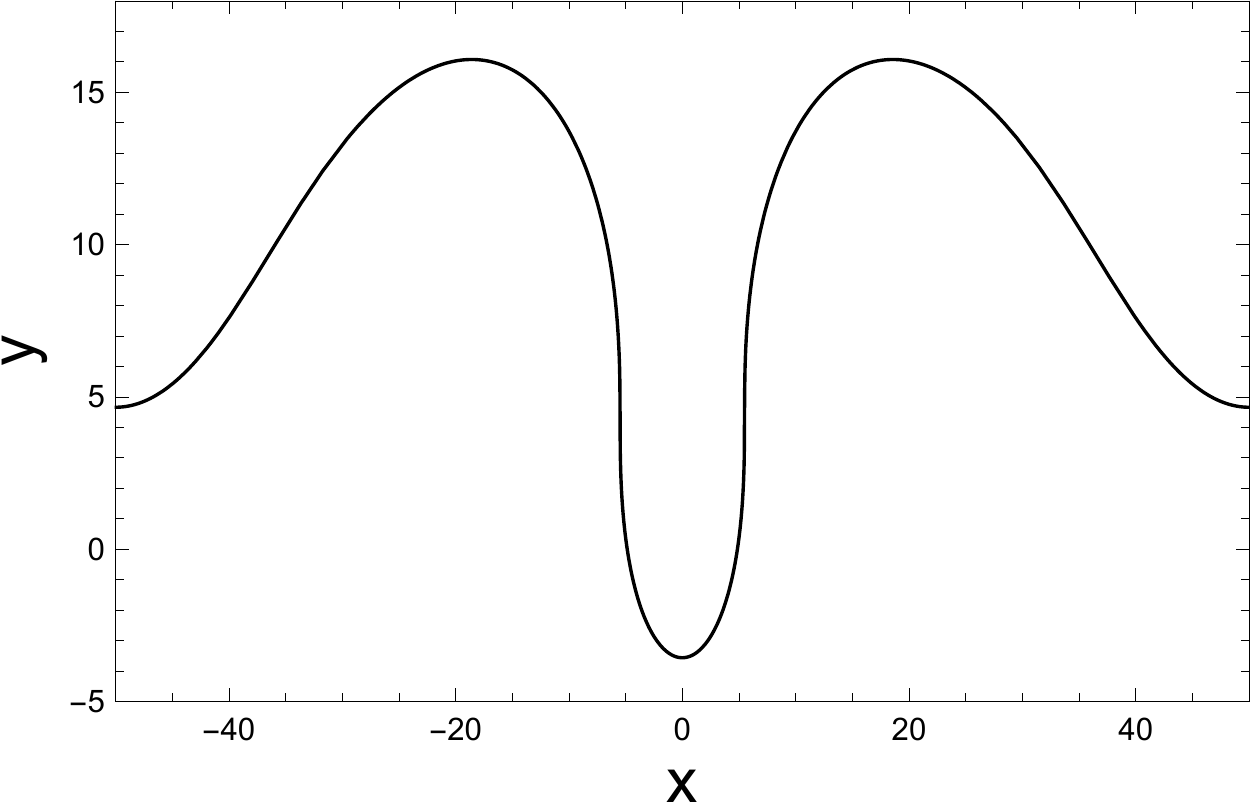}
\caption{Transient shape of the interface as predicted by \eqref{eq:velo} starting from the initial condition $-\frac{1}{10}\cos{(\pi x/50)}$ when $\Lw=0.5$ and $\Theta_i=0.8$.  The parameter $\delta=0.39$. The interface is no longer represented by a graph of a function beyond the time of this snapshot.}
  \label{fig:4}
\end{figure}

Starting from the appropriate initial data, we found traveling wave solutions for \eqref{eq:velo} and \eqref{eq:106}, whenever possible. We observed that, near the stability threshold, the interface profiles and velocities of propagation for both equatons are close to each other as shown in Fig. \ref{fig:1}. The differences increase once the relative distance from the threshold---as measured by $\delta$ defined in  \eqref{eq:delta}---becomes larger (Fig. \ref{fig:2}). Further increase in $\delta$ leads to qualitatively dissimilar solutions of the two equations. In particular, as Fig. \ref{fig:3} demonstrates, when $\delta=0.39$, the solution to \eqref{eq:106} corresponds to an interface traveling with a constant velocity. On the other hand, the solution to \eqref{eq:velo} cannot be continued beyond the transient stage as it develops vertical facets so that it can no longer be represented as a graph of a function as can be seen in Fig. \ref{fig:4}.

\subsection{Comparison  between solutions of the models \eqref{eq:velo} and \eqref{eq:ksa}}
In order to study a more complex behavior of the combustion interface, when solutions of \eqref{eq:velo} are no longer represented by traveling waves or when they fail to be represented by the graph of a function, we slightly modified our approach.
    \begin{figure}[H]
\centerline{
\includegraphics[height=2in]{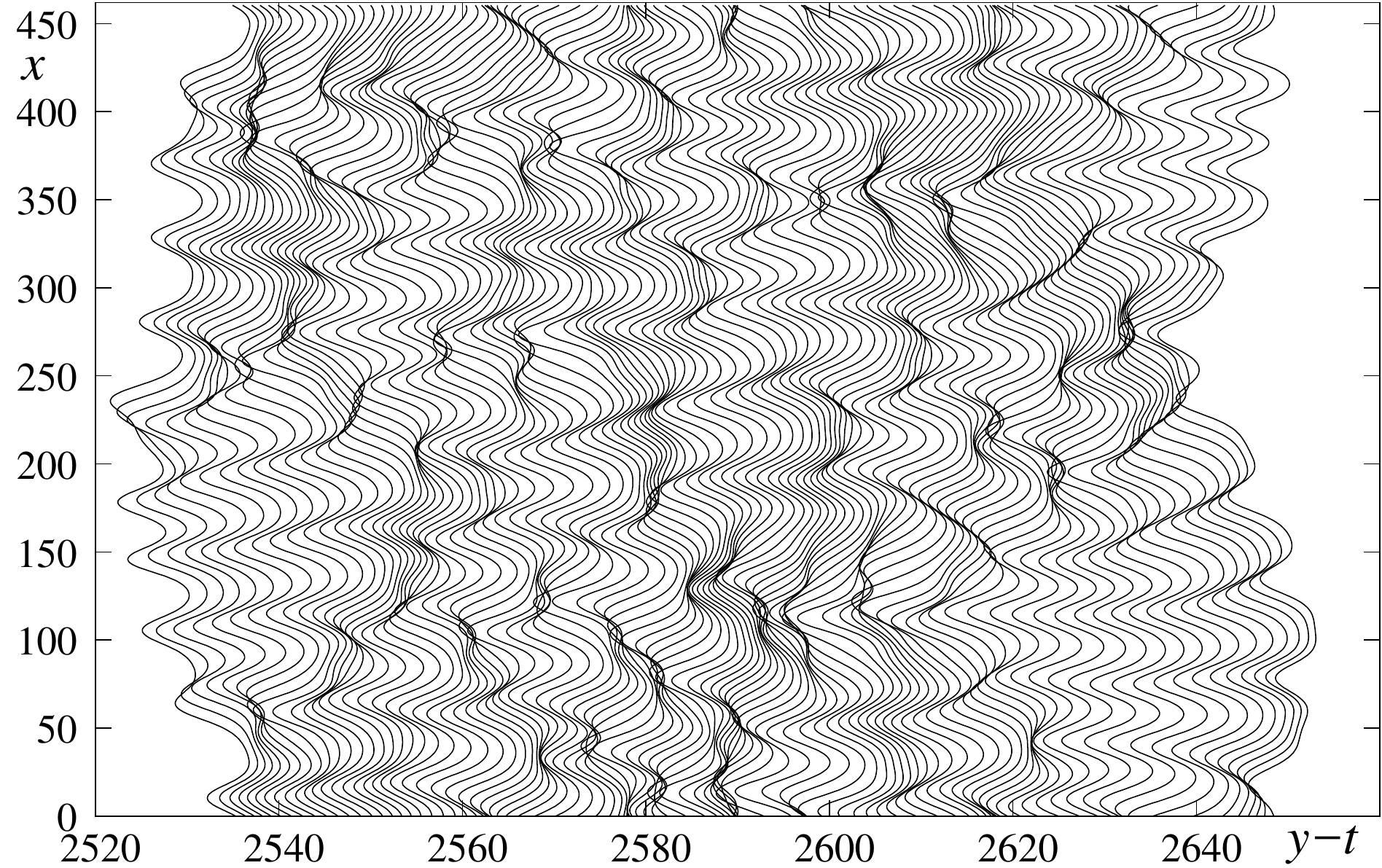}}
\caption{Numerical solution of \eqref{eq:velo} when $\Lw=0.65$ and $\Theta_i=0.75$. The parameter $\delta=0.163$. The coefficients $\gamma_1=0.914,\;\gamma_2=1.254,\;\gamma_3=21.676,\;\gamma_4=24.536$ and $0<x<450$. Shown are flame interface configurations at several consecutive equidistant instants of time ($80,000<t<100,000$). The average flame speed is $V=1.033$.}
\label{fig:110}
\end{figure}
 We expressed the interface in terms of its current arc-length variable and solved the resulting problem in a strip-like domain under the periodic boundary conditions. Here we used our own finite difference code, following the setup outlined in Appendix B. In these simulations the domain width was chosen to be long enough to support a large number of unstable modes.  We observed that in case of moderate deviations of the Lewis number from its critical value, the solutions  of \eqref{eq:velo} are relatively well behaved even in large domains and, as expected for this kind of equations \cite{gr_rev}, produce solutions with chaotic behavior, see Fig. \ref{fig:110}.

\begin{figure}[H]
\centerline{
\includegraphics[height=2in]{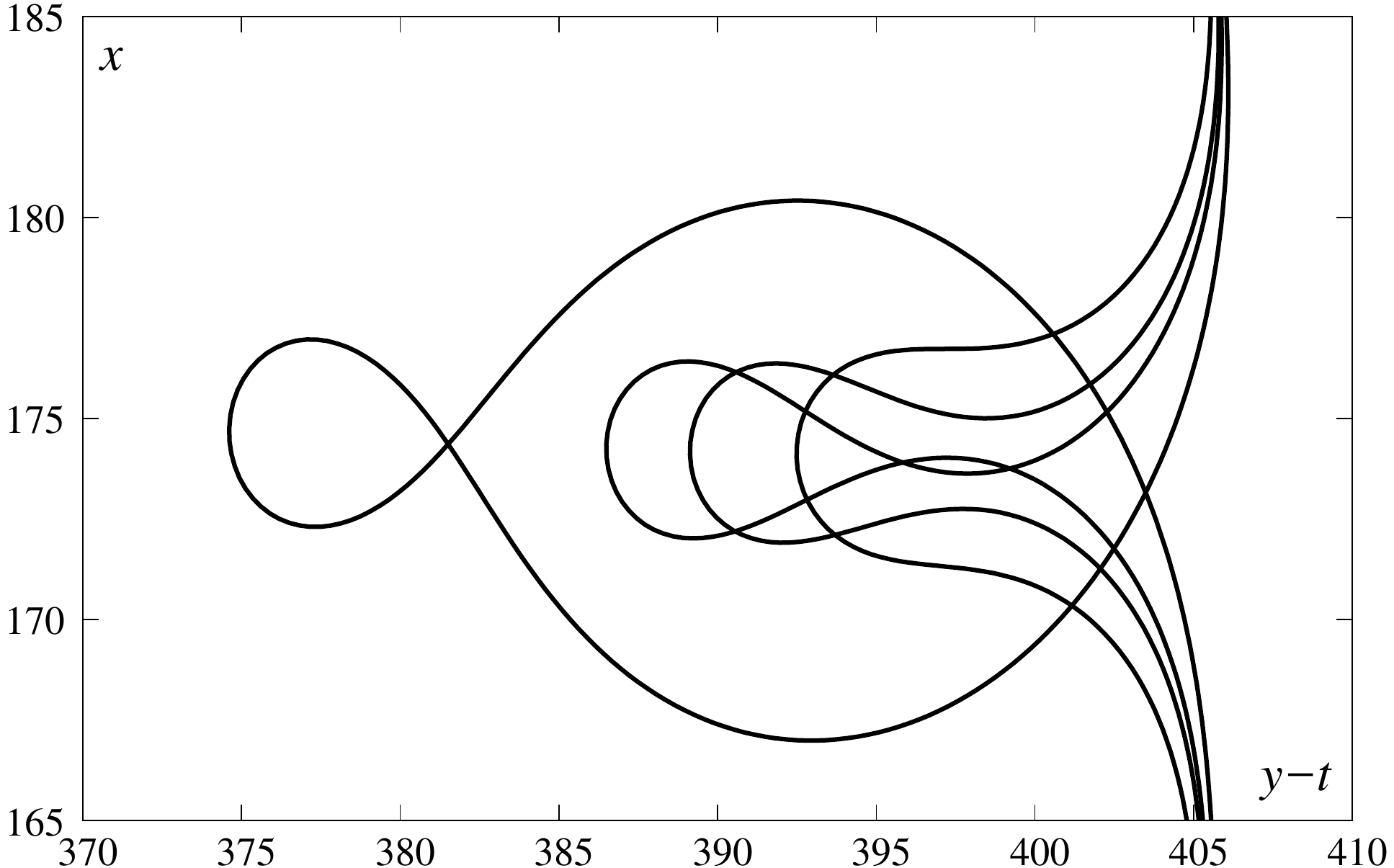}}
\caption{Numerical solution of \eqref{eq:velo} when $\Lw=0.6$ and $\Theta_i=0.75$. The parameter $\delta=0.228$. Here $0<x<500$, but is the zoom over the interval, $165<x<185,\;t=11,914;\;11,916;\;11,918;\;
11,924$.}
\label{fig:cross}
\end{figure}

However, when a deviation of the Lewis number from its critical value increases, the evolution of an initially almost planar interface eventually leads to non-physical self-intersections (Fig. \ref{fig:cross}).  We note that the results of simulations depicted on Fig. \ref{fig:cross} may only approximate the solution of the original problem  \eqref{eq:1}--\eqref{eq:3} up to the time when the interface intersects itself as the assumptions that enabled asymptotic reduction of \eqref{eq:1}--\eqref{eq:3} to \eqref{eq:velo} are no longer valid at that moment of time.

\begin{figure}[H]
\centerline{
\includegraphics[height=2in]{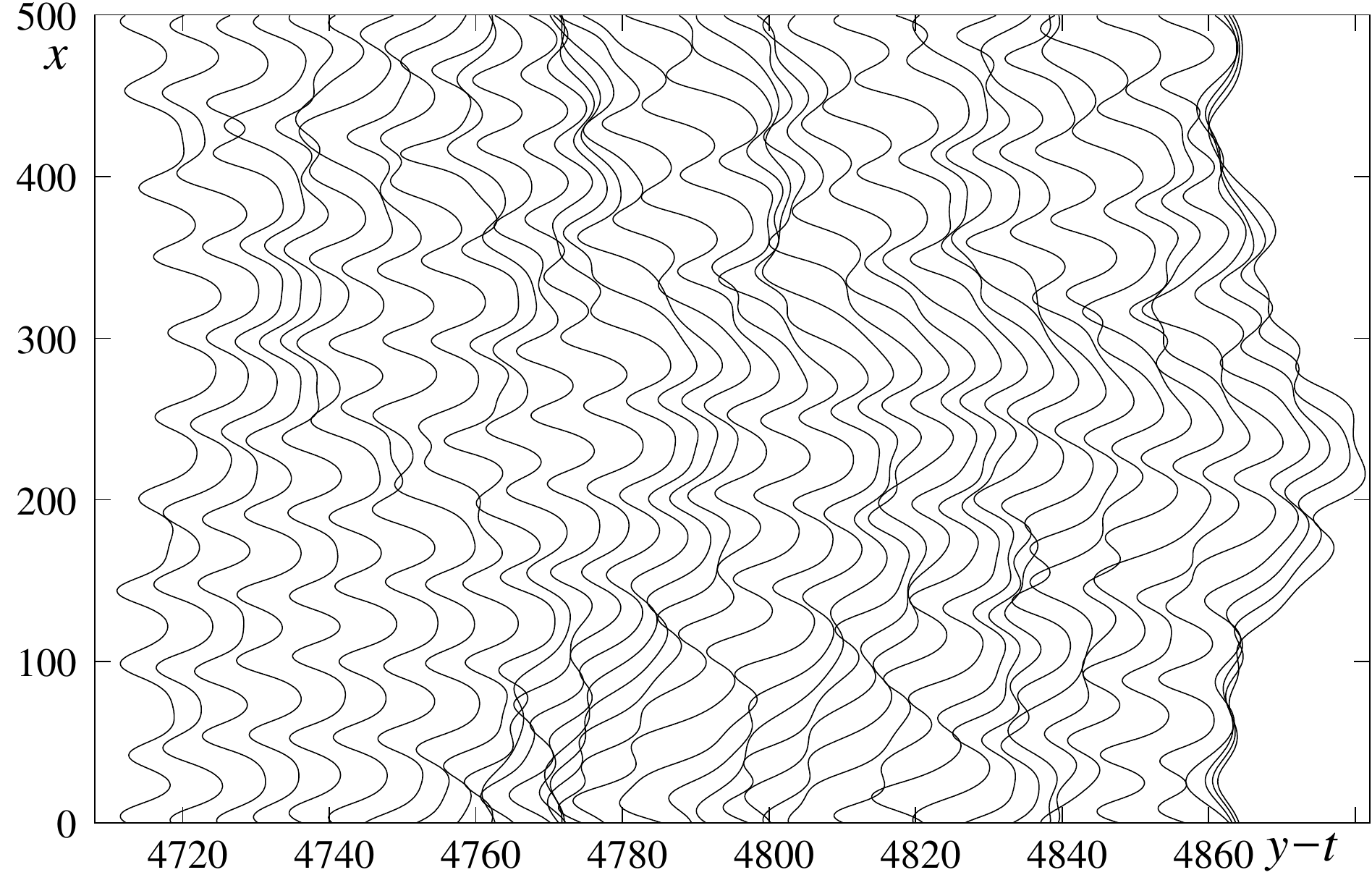}}
\caption{Numerical solution of \eqref{eq:ksa} when $\Lw=0.6$ and $\Theta_i=0.75$. The parameter $\delta=0.228$. Here $\gamma_1=1.389,\;\gamma_4=42.518$ and $0<x<500,\;1400<t<1600$. The average flame speed is $V=1.0422$. Note that $\eps=\sqrt{\gamma_1/\gamma_4}=0.18\ll1$.}
\label{long52}
\end{figure}

We note that solutions of equation \eqref{eq:ksa} appears to be free from self-intersections for any combination of $\Theta_i$ and $\Lw$ that we considered. For instance, as shown in Fig. \ref{long52}, the interface dynamics governed by \eqref{eq:ksa} does not exhibit self-intersections for the same set of parameters as those used to obtain Fig. \ref{fig:cross}. The equation \eqref{eq:ksa} preserves the spatial invariance of \eqref{eq:velo}  and elucidates the geometrical nature of the second and fourth derivatives in \eqref{eq:ks}.

\section{Concluding remarks}
\tb{In this paper, we studied an ignition-temperature, first-order reaction model of thermo-diffusive combustion that describes dynamics of thick flames
in two spatial dimensions. This model admits a planar traveling interface solution that is cellularly unstable in certain parameter regimes, when an initially flat interface develops complicated spato-temporal structures in the course of the evolution.  The interface can be identified with the level set of the solution of the combustion model that corresponds to the ignition temperature. Here we assume that an interface is a sufficiently smooth curve with small curvature that evolves in a two-dimensional space. We derive a strongly nonlinear equation for the dynamics of an interface that relates the normal velocity of the interface to its curvature, the derivatives of curvature with respect to arc-length, as well as the physical parameters of the problem (e.g., the Lewis number and the ignition temperature).  This equation represents a most general asymptotic model for the interfacial dynamics in thermo-diffusive combustion. Consequently, all other asymptotic reductions of the model \eqref{eq:1}-\eqref{eq:3} can be derived from our reduced asymptotic model by introducing appropriate rescalings. We discuss the range of applicability of our model and present its various simplifications in different parameter regimes. We also present the results of numerical simulations that demonstrate rich dynamics of the asymptotic model derived in this work.}

\section*{Acknowledgments}  The work of N.K.,  P.V.G.,  L.K. and G.I.S.  was
supported, in part, by the US-Israel Binational Science Foundation under the grant 2012057.
The work of D.G. was supported in part by the NSF grant DMS-1615952.
The work of P.V.G was also supported by the grant 317882 from Simons Foundation.
The work of  L.K. and G.I.S was also supported  by the Israel Science Foundation (Grant 335/13). 
The computational component of this work was also supported by the Ohio Supercomputer Center  grant PBS0293-1.
P.V.G. also would like to thank John Coleman for creating an excellent work environment.

\section*{Appendix A}

In this appendix we state several standard formulas from differential geometry of planar evolving curves. Details can be found in any introductory text on the subject, e.g., \cite{Gurtin}, see also Appendix in \cite{pre}.

Let $\mathcal{C}$ be a simple, smooth evolving curve with bounded curvature, $k$.  This curve can be represented by its position vector $\mathbf{R}(s,t)$, where $s$ is the arc-length parameter of the curve and $t$ is time. Then a fixed point in the vicinity of the curve is defined by position vector 
\begin{equation}
\mathbf{r}(n,s,t)=\mathbf{R}(s,t)+n\mathbf{N}(s,t), \nonumber
\end{equation}  
where $\mathbf{N}$ is a unit normal vector to $\mathcal{C}$ and $n$ is the distance from the curve to the point.
The unit tangent vector $\mathbf{T}$ is  given by 
\begin{equation}
\mathbf{T}=\dfrac{\partial \mathbf{R}}{\partial s}.\nonumber
\end{equation}
In what follows we will use $n,s,t$ as local coordinates in the vicinity of curve $\mathcal{C}$.

In the setting given above  the Frenet relations take the form
\begin{equation}
\begin{array}{rcl}
\dfrac{\partial}{\partial s}\mathbf{N}&=&k \mathbf{T}, \\ \\
\dfrac{\partial}{\partial s}\mathbf{T}&=&-k \mathbf{N}.
\end{array}
\end{equation}
As follows from Frenet formulas  the curvature of $\mathcal{C}$ is given by
\begin{equation}
k=\mathbf{T}\cdot\dfrac{\partial}{\partial s}\mathbf{N}=-\mathbf{N}\cdot\dfrac{\partial}{\partial s}\mathbf{T}.\nonumber
\end{equation}

The spatial gradient in terms of local coordinates is defined as
\begin{equation}
\nabla=\mathbf{N}\dfrac{\partial}{\partial n}+\mathbf{T}\dfrac{1}{1+nk}\dfrac{\partial}{\partial s}.\nonumber
\end{equation}
Consequently, for any scalar function $\phi(n,s,t)$, the Laplacian operator is given by
\begin{equation}
\Delta \phi=\dfrac{\partial^2 \phi}{\partial n^2} + \dfrac{k}{1+nk}\dfrac{\partial \phi}{\partial n} + \dfrac{1}{1+nk}\dfrac{\partial}{\partial s}\left(\dfrac{1}{1+nk}\dfrac{\partial \phi}{\partial s}\right).\label{space}
\end{equation}

The velocity of the curve can be expressed in terms of its components, the normal velocity, $V_n$ and the tangential velocity $V_\perp$.  The components are defined as
\begin{equation}
V_\perp=-\dfrac{\partial\mathbf{R}}{\partial t}\cdot\mathbf{T},
\end{equation}
and
\begin{equation}
V_n=\dfrac{\partial\mathbf{R}}{\partial t}\cdot\mathbf{N}.
\end{equation}

For any scalar function $\phi(n,s,t)$ the material time derivative reads
\begin{equation}
\dfrac{\mathrm{D}\phi}{\mathrm{D}t}= - V_n\dfrac{\partial\phi}{\partial n}+\mathcal{L}_t\phi,\label{time}
\end{equation}
where the Lagrangian time derivative, $\mathcal{L}_t$ defined as
\begin{equation}
\mathcal{L}_t=\dfrac{\partial}{\partial t} + V_\perp\dfrac{\partial}{\partial s}.
\end{equation}
Normal and tangent velocity of the curve are related via transport identity 
\begin{equation}
 \dfrac{\partial}{\partial s}V_\perp=k V_n.\label{trans}
\end{equation}
Finally, the  Lagrangian time derivative of the curvature, $\mathcal{L}_t\kappa$, can be expressed as follows
\begin{equation}
\mathcal{L}_t\kappa=-\dfrac{\partial^2}{\partial s^2}V_{n}-\kappa^2 V_n.\label{transport}
\end{equation}

\section*{Appendix B}
In this section we outline our setup for the numerical simulations of \eqref{eq:velo} and \eqref{eq:ksa} considered in a strip-like domain.

For numerical simulations of  \eqref{eq:velo} it is more convenient to use unscaled quantities and introduce the small parameter via the initial conditions.
It is also most convenient to represent a point located on the interface by using its current position in Cartesian coordinates.
Thus, we set the function
\begin{eqnarray}
\mathbf{R}(s,t)=x(s,t){\bf i}+y(s,t){\bf j},
\end{eqnarray}
where
\begin{eqnarray}\label{eq:ab1}
\left(\frac{\partial{x}}{\partial{s}}\right)^2+
\left(\frac{\partial{y}}{\partial{s}}\right)^2=1,
\end{eqnarray}
to represent the interface at the time $t>0$. In this setting the curvature of the interface is given by
\begin{eqnarray}
\label{eq:ab2}
\kappa=\frac{\partial{^2y}}{\partial{s^2}}\frac{\partial{x}}{\partial{s}}-
\frac{\partial{^2x}}{\partial{s^2}}\frac{\partial{y}}{\partial{s}}.
\end{eqnarray}
The normal and tangential velocity $V_n$ and $V_{\perp}$ are connected with $\mathbf{R}$ via the following relations
\begin{eqnarray} \label{eq:ab3}
&& \frac{\partial{x}}{\partial{t}}+V_{\perp}\frac{\partial{x}}{\partial{s}}=
V_n\frac{\partial{y}}{\partial{s}}, \nonumber \\
&& \frac{\partial{y}}{\partial{t}}+V_{\perp}\frac{\partial{y}}{\partial{s}}=
-V_n\frac{\partial{x}}{\partial{s}}.
\end{eqnarray}
Finally, 
 the rate of the arc-length stretch is given by
\begin{eqnarray}\label{eq:ab4}
\frac{dL}{dt}=\int_0^L \kappa V_n d\hat{s},
\end{eqnarray}
Consequently, for a given time $t$, the range of arc-length is $0<s<L(t)$.

Substituting the equations \eqref{eq:ab1}-\eqref{eq:ab3} and \eqref{trans} into \eqref{eq:velo} or \eqref{eq:ksa} and using \eqref{eq:ab4} to
define computational domain, we end up with a closed system of PDEs for $x$ and $y$ which should be complemented with boundary and initial conditions.
For simplicity we assume periodic boundary conditions and initial conditions
\begin{eqnarray}
\frac{dx_0}{ds}=-\sin{\Psi},\quad \frac{dy_0}{ds}=\cos{\Psi},
\end{eqnarray}
where
\begin{eqnarray}
\Psi(s)=\frac{\pi}{2}+a\cos{\left(\frac{2\pi s}{L_0}\right)}+
b\sin{\left(\frac{6\pi s}{L_0}\right)}.
\end{eqnarray}
In this expression, $L_0=L(0)$ is the initial arc-length with $L_0\simeq l$ (where $l$ is a length of the strip) and $a\ll 1,\;b\ll 1$.  
For computations discussed in this paper, we set $a=0.01,\;b=0.005,\;l=10\left(2\pi/k_c\right),$  where $k_c=\sqrt{\frac{\gamma_1}{2\gamma_4}}$ is the wavelength corresponding to the maximum growth rate and $\omega=\gamma_1 k^2-\gamma_4 \gamma^4$ are obtained from the linear stability analysis. As we have already mentioned above, $\gamma_1/\gamma_4$ and therefore $k_c$ are typically small, hence $l\gg1$ in our computations. 

The numerical simulations of the resulting systems of equations were performed using conventional finite difference methods.

\end{document}